\documentclass[aps,pra,twocolumn,floatfix]{revtex4-2}
\usepackage{blindtext}
\usepackage{bm}
\usepackage{braket}
\usepackage{hyperref}
\usepackage{amssymb}
\usepackage{amsmath}
\usepackage{graphicx}
\usepackage{float}
\usepackage{bbold}
\usepackage{mathtools}
\begin{document}
\title{Measurement noise susceptibility in quantum estimation}
\author{Stanisław Kurdziałek}
\email{s.kurdzialek@student.uw.edu.pl}
\affiliation{Faculty of Physics, University of Warsaw, Pasteura 5, 02-093 Warszawa, Poland}
\author{Rafał Demkowicz-Dobrza{\'n}ski}
\affiliation{Faculty of Physics, University of Warsaw, Pasteura 5, 02-093 Warszawa, Poland}
\begin{abstract}
Fisher Information is a key notion in the whole field of quantum metrology. It allows for a direct quantification of maximal achievable precision of estimation of parameters encoded in quantum states using the most general quantum measurement. It fails, however, to quantify the robustness of quantum estimation schemes against measurement imperfections, which are always present in any practical implementations. Here, we introduce a new concept of Fisher Information Measurement Noise Susceptibility  that quantifies the potential loss of Fisher Information due to small measurement disturbance. We derive an explicit formula for the quantity, and demonstrate its usefulness in analysis of paradigmatic quantum estimation schemes, including interferometry and super-resolution optical imaging.
\end{abstract}

\maketitle

\paragraph*{Introduction.} 
Noise, decoherence and implementation imperfections are the main factors hindering the transfer of quantum enhanced technologies (e.g. quantum computing and communication) from proof-of-principle experiments to real-life applications \cite{Preskill2018}.
 These issues also affect the development of quantum metrology, whose goal is to utilize sophisticated properties of light and matter to enhance sensing instruments \cite{Giovannetti2011, Demkowicz2015, Degen2017, PezzeSmerziOberthalerEtAl2018}. Quantum estimation theory \cite{Holevo1982, Braunstein1994} laid theoretical grounds for present-day quantum metrology---one of its greatest achievements is identification of protocols that perform optimally in the presence of noise \cite{Escher2011, Demkowicz2012, Demkowicz2017, Zhou2018, Zhou2021}. 

One of the key elements affecting the precision of metrological protocols is the imperfect realization of the final measurement step, where information is being extracted from quantum sensors. In order to asses the effect of imperfect measurement implementation, the standard route is to characterize the type of noise present, e.g. detector dark counts, measurement outputs cross-talks, etc. and then analyze its impact on the relevant figures-of-merit \cite{Datta2011}. A more systematic and general study of the effect of readout noise on the measurement precision is provided in Ref.~\cite{Len2021}.

Still, from a fundamental point of view, it would be much more advantageous to be able to determine noise robustness of a given measurement scheme without specifying the actual form of the noise. A similar motivation lays behind measurement robustness considerations that can be found in the context of other quantum information tasks, see e.g. \cite{Skrzypczyk2019, Oszmaniec2019}, but have never been applied to quantum estimation theory. 
In this paper we focus on the most common figure of merit in quantum estimation theory---the Fisher Information---and propose a quantity, Fisher Information Measurement Noise Susceptibility (FI MeNoS), which characterizes the maximal relative decrease of FI due to small measurement disturbance of the most general type. This quantity allows us to obtain a deep insight into fundamental noise-robustness properties of different measurement schemes without assuming any particular noise form. We illustrate fruitfulness of this approach by analysing paradigmatic quantum enhanced metrological schemes including interferometry and superresolution imaging.

\paragraph*{Quantum Estimation Theory Preliminaries.}
In a paradigmatic quantum estimation scenario, a continuous parameter  $\theta$ is encoded in a state $\rho_\theta$ of a probe system with associated  Hilbert space $\mathcal{H}_\textrm{S}$. In order to describe the process of extraction of information on the parameter $\theta$ from the state in full generality, one considers an external measuring device whose Hilbert space is $\mathcal{H}_\textrm{M}$. The device is initialized in a pure state $\ket{0}_\textrm{M}$ and the generalized measurement of $\rho_\theta$ consists of two stages: (i) interaction between S and M described by a unitary operation $U_\textrm{SM}$; (ii) projective measurement of the post-evolution state of M, which returns an outcome $i \in \{ 1,2,...,K\}$ with a probability
\begin{equation}
\label{eq:prob}
    p_\theta(i) =  \textrm{Tr} \left( \rho_\theta M_i \right),\quad     M_i = \prescript{}{\textrm{M}}{\braket{0|U^\dagger_\textrm{SM}|i}_\textrm{M}\!\!} \braket{i|U_\textrm{SM}|0}_\textrm{M}
\end{equation}
where $M_i$ are effective measurement operators acting on S---note that scalar products in the above formula are partial, they act on subsystem M only, leaving part S intact. 
 A set $\bm{M} = \{M_i\}_i$ is called a positive operator-valued measure (POVM), where $M_i$ satisfy (i) $\sum_i M_i = \mathbb{1}$, (ii) $M_i \ge 0$.
 The set of all POVMs will be denoted as $\mathcal{M}$, so we will write $\bm{M} \in \mathcal{M}$.
 Different choices of $U_\textrm{SM}$ lead to different POVMs, the number of possible outcomes is $K = \textrm{dim} \mathcal{H}_\textrm{M}$. Each POVM can be physically implemented with the help of an appropriate choice of $\mathcal{H}_\textrm{M}$ and $U_\textrm{SM}$. The projective measurement in a basis $\ket{i}$ of $\mathcal{H}_\textrm{S}$ corresponds to $\bm{M}$ with $M_i = \ket{i}\!\! \bra{i}$.
 
When $N$ copies of $\rho_\theta$ are measured independently with the same POVM $\bm{M}$
this leads to $N$ i.i.d. random variables 
sampled from $p_\theta(i)$.  
According to the Cram\'er-Rao bound  (CRB) \cite{Holevo1982, Braunstein1994}, the mean squared error (MSE) of any (locally) unbiased  estimator  $\tilde \theta$, that estimates $\theta$ based on this data,  will be lower bounded as
\begin{equation}
\label{CCRB}
    \Delta^2 \tilde \theta \ge \frac{1}{N F_\textrm{C} }, \quad F_\textrm{C} = \sum_i p_i l_i^2
\end{equation}
where $F_C$ is the classical Fisher Information (CFI), $p_i=p_\theta(i)$ and  $l_i = \partial_\theta \log p_\theta(i)$ is the  logarithmic derivative of $p_\theta(i)$. 
Intuitively, the CFI quantifies how sensitive is $p_\theta(i)$ to the change of $\theta$---the larger $l_i^2$, the greater the CFI. The CRB is tight---it is always possible to find a locally unbiased estimator whose MSE saturates \eqref{CCRB}, and for $N \rightarrow \infty$ one can construct a globally unbiased CRB-saturating estimator \cite{vanderVaart1998}.

For a fixed quantum state $\rho_\theta$, the CFI depends only on the measurement $\bm{M}$, and in order to highlight this we will denote it
as $F_C[\bm{M}]$. Combining \eqref{CCRB} with \eqref{eq:prob} the explicit form of CFI reads:
\begin{equation}
\label{CFIdef}
    F_\textrm{C} \left[ \bm{M}\right] =  \sum_i \textrm{Tr}(\rho_\theta M_i) l_i^2,\quad l_i = \frac{\textrm{Tr} (\dot \rho_\theta M_i)}{\textrm{Tr} ( \rho_\theta M_i)},
\end{equation}
where dot denotes the derivative over $\theta$. 
 It is natural to ask, what is the greatest possible CFI for a given $\rho_\theta$---the answer is given by Quantum Fisher Information (QFI) \cite{Holevo1982, Braunstein1994}, which is the maximum of the CFI over all POVMs $\bm{M}$, and can be computed as
\begin{equation}
    F_\textrm{Q} = \max_{\bm{M} \in \mathcal{M}} F_{\textrm{C}}[\bm{M}] = \textrm{Tr} \left( \rho_\theta \Lambda_\theta^2 \right),
\end{equation}
where $\Lambda_\theta$ is the symmetric logarithmic derivative matrix defined by the equation
$
\partial_\theta \rho_\theta = \frac{1}{2} \left( \rho_\theta \Lambda_\theta + \Lambda_\theta \rho_\theta \right).
$
 For a given $\rho_\theta$, and arbitrary $\bm{M}$, the MSE of any locally unbiased estimator of $\theta$ is lower-bounded by Quantum Cram\'er-Rao bound (QCRB), which is similar to \eqref{CCRB}, but $F_\textrm{C}$ is replaced with $F_\textrm{Q}$. The projective measurement on eigenstates of $\Lambda_\theta$ is always QCRB-saturating (its CFI is equal to QFI), but sometimes there are many different QCRB-sat. measurements, see \cite{Zhou2020} and Appendix \ref{AppD} for a detailed discussion.

\paragraph*{Fisher Information Measurement Noise Susceptibility.}

Let us assume, that due to a small disturbance, $\bm{M}$ changes to $ \bm{ \tilde M} = (1-\epsilon) \bm{M} + \epsilon \bm{N}$ (summation of two POVMs is done element-wise), which can be viewed as  replacement of desired POVM $\bm{M}$ with an unwanted one $\bm{N}$ with a probability $\epsilon \ll 1$. This type of noise may be caused by inaccurate initialization of a measuring device M in a mixed state $(1-\epsilon)\ket{0}\!\!\bra{0} + \epsilon \rho_\textrm{M}'$ instead of $\ket{0}\!\!\bra{0}$, or it may be the result of other small imperfections, such as signal losses, dark counts, cross-talks etc. (see Appendices  A and B).

The measurement noise affects the CFI, effect of which we quantify using
\begin{equation}
\label{chiMN}
\chi[\bm{M}, \bm{N}] = \lim_{\epsilon \rightarrow 0} \frac{F_\textrm{C} \left[ \bm{M} \right] -  F_\textrm{C} [ (1-\epsilon)\bm{M} + \epsilon \bm{N}]}{\epsilon \cdot  F_\textrm{C} [ \bm{M}]},
\end{equation}
which can be understood as the relative \emph{decrease} of CFI under infinitesimally added noise $\bm{N}$---the effect of $\epsilon$-noise results in $F_C$ in CRB, Eq.~\ref{CCRB}, being replaced with $F_C[\bm{M}](1 - \epsilon \chi[\bm{M},\bm{N}])$.  
 After inserting \eqref{CFIdef} into \eqref{chiMN}, we obtain, after straightforward calculations,
\begin{equation}
\label{chiMN2}
    \chi \left[ \bm{M}, \bm{N} \right] = 1+ F_\textrm{C}[\bm{M}]^{-1} G[\bm{N}],
\end{equation}
where 
\begin{equation}
\label{Gdef}
G[\bm{N}] = \sum_i \textrm{Tr} (A_i N_i),\quad A_i=l_i^2 \rho_\theta - 2 l_i \dot \rho_\theta.
\end{equation} 

To get an intuition regarding this quantity, consider a simple example where  $\bm{M}=(0,M_2,...,M_K)$ and  $\bm{N}=(\mathbb{1},0...,0)$, so noise only activates a non-informative measurement outcome $1$. Then,  $G[\bm{N}]$=0 and hence $\chi \left[\bm{M}, \bm{N}\right]=1$, which means that the  relative decrease of the CFI is equal to the probability of obtaining a useless, noisy result. 
Clearly, the decrease of CFI will be more substantial, when the disturbance affects the statistics of informative outcomes, and noise cannot be separated from the signal easily.
Our goal is to figure out, what is the maximal shrinkage rate of the CFI caused by an infinitesimal measurement noise described by an arbitrary POVM $\bm{N}$. The answer is given by a quantity
\begin{equation}
\label{chiM}
      \chi \left[\bm{M} \right] = \underset{\bm{N} \in \mathcal{M}}{\textrm{max}}\  \chi [\bm{M}, \bm{N}],
\end{equation}
which we call FI MeNoS because it tells us, how \textit{susceptible} is the CFI to small disturbances of the measurement. Note that the larger $\chi$ implies potentially stronger decrease of CFI as a result of measurement disturbance, so strictly speaking this is a
\emph{negative} susceptibility (cf. `menos' in Spanish).
Notably, it does not depend on $ \bm{N} $---this allows us to compare the robustness against noise of different measurements without invoking any specific noise model.

\paragraph*{Explicit formula for FI MeNoS.}
\begin{figure*}[t]
\centering
\includegraphics[width=0.95\textwidth]{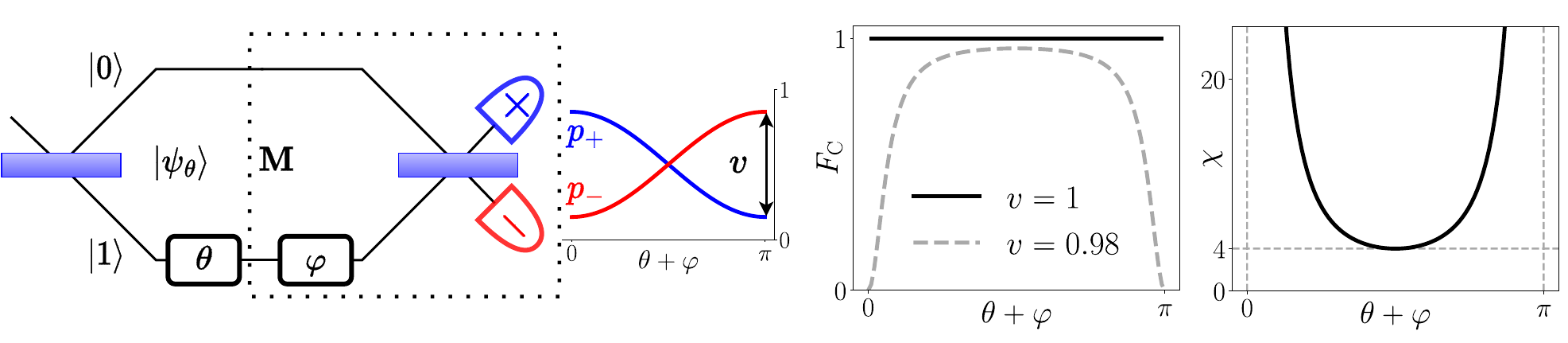}
\caption{Phase $\theta$ is measured using Mach-Zender interferometer, $\varphi$ is an extra, well-controlled phase. The resulting CFI ($F_\textrm{C}$) does not depend on $\varphi$ when $v=1$, which is never achieved in practive. For any smaller visibility (e.g. $v=0.98$), the CFI is maximal for $\theta+\varphi=\pi/2$, and vanishes for $\theta+\varphi \in \{0,\pi\}$. We can deduce that $\theta+\varphi=\pi/2$ is the optimal working point without assuming non-unit visibility because FI MeNoS ($\chi$) is minimal there, so the precision of the estimation of $\theta$ is the least vulnerable to a general measurement noise.  }
\label{Fig:interferometer}
\end{figure*}  
We now present an explicit solution to the maximization problem from \eqref{chiM}, which, according to \eqref{chiMN2}, boils down to finding the maximum of $G[\bm{N}]$. Without loss of generality, we can relabel the elements of POVM $\bm{M}$ such that logarithmic derivatives satisfy $l_1 \le l_2 \le ... \le 
l_K$. Let $\bm{N}=(N_1,...,N_i,...,N_K)$ be an arbitrary POVM while  $\Gamma_{i}^1 (\bm{N})=(N_1+N_i,...,0,...,N_K)$ and $\Gamma_{i}^K (\bm{N})=(N_1,...,0,...,N_K+N_i)$ be two POVMs constructed from $\bm{N}$ with zeros at $i$-th positions, $i \in \{2, \dots, K-1\}$.  Using \eqref{Gdef}, we obtain:
\begin{align}
\label{G1}
   G[\Gamma^1_i \left( \bm{N} \right)] - G[\bm{N}] =f_i(l_1) - f_i(l_i), \\
   \label{G2}
   G[\Gamma^K_i \left( \bm{N} \right)] - G[\bm{N}] =f_i(l_k) - f_i(l_i),
\end{align}
  where $f_i(x)=x^2 \textrm{Tr}(\rho_\theta N_i) - 2 x \textrm{Tr} (\dot \rho_\theta N_i)$ is a convex quadratic function, and therefore  $ f_i(l_1) \ge f_i(l_i) \lor f_i(l_K) \ge f_i(l_i) $, so from \eqref{G1} and \eqref{G2}, $G[\Gamma^1_i \left( \bm{N} \right)] \ge G[\bm{N}] \lor G[\Gamma^K_i \left( \bm{N} \right)] \ge  G[\bm{N}]$. Therefore, for each $\bm{N}$ and $i$, there is $j_i \in \{1,K\}$ such that $G[\Gamma^{j_i}_i(\bm{N})] \ge G[\bm{N}]$, so we can choose $i_2,...,i_{K-1} \in \{1,K\}$ such that $\bm{\tilde N} = \Gamma_2^{j_2} \circ... \circ \Gamma_{K-1}^{j_{K-1}}(\bm{N}) $ satisfies $G[ \bm{\tilde N}] \ge G[\bm{N}] $, and from the construction of $\bm{\tilde N}$, $\tilde N_i =0 $ for $i \in \{2,...,K-1\}$. It means, that for arbitrary $\bm{N}$, it is possible to construct $\bm{\tilde N} = (\tilde N_1,0,...,0,\mathbb{1} - \tilde N_1)$ satisfying $\chi[\bm{M}, \bm{\tilde N}] \ge \chi[\bm{M}, \bm{N}]$. Hence, the worst-case scenario noise will affect only the outcomes with the smallest and the largest logarithmic derivatives. This observation allows to perform the maximization from \eqref{chiM} analytically, (see Appendix \ref{AppC}), and obtain the general expression for  the FI MeNoS:
 \begin{equation}
 \label{chian}
     \chi [\bm{M}] = 1 + \frac{1}{2 F_\textrm{C}[\bm{M}]} \left( l_1^2+l_K^2 + \|A_1-A_K \|_1 \right),
 \end{equation}
 where $\| A \|_1 = \textrm{Tr}\sqrt{A A^\dagger}$ is the trace norm. Notice, that $\chi$ depends on $F_\textrm{C}$, $\rho_\theta$, $\dot \rho_\theta$ and the extremal logarithmic derivatives only. 
When an outcome $i$ has a vanishing probability, $p_i \rightarrow 0$ but its contribution to the CFI, $p_i l^2_i$ remains finite and non-zero, then $l_i^2 \rightarrow \infty$, which implies that either $l_1^2$ or $l_k^2$ diverges, and hence, $\chi$ diverges as well according to \eqref{chian}. This reflects the fact, that the contribution to the CFI resulting from an outcome with a very low probability  may be completely washed out by a very small measurement noise, and the chosen measurement is not likely to be practical.

 When there are many measurements which lead to the same CFI, the FI MeNoS may help to judge, which one is more robust and hence more suitable for practical purposes---the one with lower $\chi$. 
It is especially interesting to find the minimum of $\chi \left[ \bm{M} \right]$ over all  QCRB-sat. measurements, 
\begin{equation}
    \chi_\textrm{Q} = \underset{\left\{\bm{M} \in \mathcal{M},  F_\textrm{C}\left[ \bm{M} \right] = F_\textrm{Q}\right\}}{\textrm{min}} \chi \left[ \bm{M} \right],
\end{equation}
since the corresponding measurement $\bm{M}$ should be regarded as the most robust among the most informative measurements.
This task is tractable  thanks to the exact formula \eqref{chian}---we demonstrate exemplary solutions to this problem in the next two paragraphs.

\paragraph*{Pure state models.}

 Let us start with a simple, yet important case when $\rho_\theta$ is pure, $\rho_\theta = \ket{\psi_\theta}\!\! \bra{\psi_\theta}$. We focus on the local estimation paradigm and assume $\theta$ is close to some known parameter value $\theta_0$. For any $\theta_0$ it is possible to fix orthonormal vectors $\ket{0}, \ket{1} \in \textrm{span}\left( \ket{\psi_{\theta_0}}, \ket{\dot \psi_{\theta_0}} \right)$ such that $\rho_{\theta_0} = \ket{+}\!\!\bra{+}$, $\dot \rho_{\theta_0} = \frac{1}{2} \sqrt{F_\textrm{Q}} \sigma_y$, where $\ket{+} = \frac{1}{\sqrt{2}} \left( \ket{0} + \ket{1} \right)$, $\sigma_y$ is a Pauli matrix, $F_\textrm{Q}  
 $ is the QFI. Then, the measurement $\bm{M}$ is QCRB-sat. iff all its elements are of the form $M_i = \lambda_i \ket{\phi_{i}} \bra{\phi_{i}}$, where $\ket{\phi_i} = \frac{1}{\sqrt{2}} \left( \ket{0} + e^{i \varphi_i} \ket{1} \right)$ (see Appendix  \ref{AppE1}). As we prove in Appendix \ref{AppE1}, $\chi[\bm{M}] \ge 4 F_\textrm{Q}$ for all such measurements, and the inequality is only saturated for the projective measurement on the eigenstates of $\sigma_y$. Notice, that we used a qubit subspace to describe the evolution of any pure state locally even though $\mathcal{H}_\textrm{S}$ can be arbitrarily large.

This parametrization allows to represent any pure state problem as a phase estimation in a Mach-Zender interferometer with a single photon input. When the phase $\theta$ between upper ($\ket{0}$) and lower ($\ket{1}$) arm is acquired, then the photon state is $\ket{\psi_\theta} = \frac{1}{\sqrt{2}} \left( \ket{0} + e^{i \theta} \ket{1} \right)$. After fixing $\theta_0=0$, our problem reduces to the one already defined with $F_\textrm{Q}=1$. All QCRB-sat. projective measurements can be implemented with the help of two single-photon detectors followed by a beam-splitter, and a well-controlled phase difference between two arms, $\varphi$ (see Fig.~\ref{Fig:interferometer}). The upper and lower  detectors click with probabilities $p_+$ and $p_-$ respectively, where $p_\pm  = \frac{1}{2} \left[ 1 \pm \cos(\theta+\varphi) \right]$. Straightforwad calculations confirm, that $F_\textrm{C}[p_\pm]=1$ independently of $\varphi$ \footnote{Strictly speaking, we cannot use \eqref{CCRB} directly for $\varphi \in \{ 0, \pi \}$ because then $p_i=0$ and $\dot p_i = 0$ for one term of the sum from \eqref{CCRB}. If we exclude such terms by convention, we obtain $F_\textrm{C}=0$, but if we calculate limits $\varphi \rightarrow 0$ or $\varphi \rightarrow \pi$, we obtain $F_\textrm{C}=1$}. However, the FI MeNoS  depends on $\varphi$, and for $\theta_0=0$ we have
\begin{equation}
\label{chiint}
    \chi(\varphi)  = 1+ \cos^{-2}(\varphi/2) + \tan^{-2}(\varphi/2),
\end{equation}
 which means that the optimal working point is at $\varphi= \pi/2$ (balanced interferometer), while our scheme is extremely sensitive to a measurement noise for $\varphi \rightarrow 0$ and $\varphi \rightarrow \pi$ (unbalanced interferometer) as in this case $\chi (\varphi) \rightarrow \infty$, see Fig.~\ref{Fig:interferometer}.
 \begin{figure*}[t]
    \centering
    \includegraphics[width=0.95\textwidth]{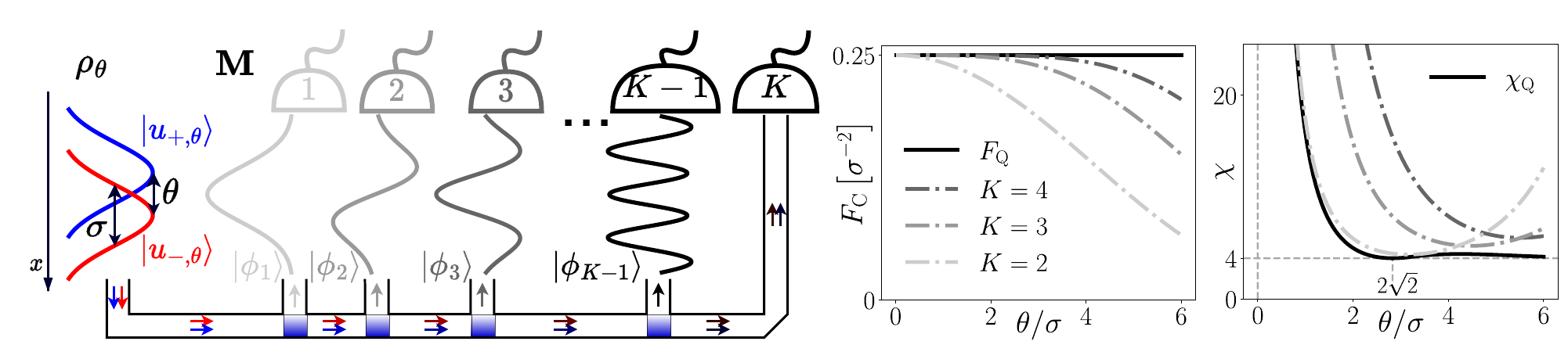}
    \caption{The single photon state ($\rho_\theta$) in an image plane is the mixture of two Gaussian wave functions of width $\sigma$. Their separation $\theta$ can be estimated  accurately even when $\theta \ll \sigma$ by measuring $\rho_\theta$ in the basis of Hermite-Gaussian modes $\ket{\phi_i}$---in practice, it is enough to extract first few ($K-1$) of these modes. Unfortunately, the presented strategy is very sensitive to measurement noise for small $\theta$, which is reflected by diverging FI MeNoS ($\chi$) for $\theta/\sigma \rightarrow 0$. No QCRB-sat. measurement is free of this issue because $\chi_\textrm{Q}$ diverges as well for $\theta \rightarrow 0$.}
    \label{Fig:superresolution}
    
\end{figure*}

 Similar conclusions follow from a standard analysis of the non-unit visibility ($v<1$) interferometer model, where the detection probabilities are $p_\pm = \frac{1}{2} \left( 1 \pm v \cos (\theta) \right)$. Then, the CFI is  maximal for $\varphi = \pi/2$, and reaches $0$ for $\varphi \in \{0,\pi\}$ even for $v$ very close to $1$, see Fig.~\ref{Fig:interferometer}). The advantage of the approach based on FI MeNoS, is that one does not need to consider any particular noise model, and it is guaranteed that the worst case scenario has been taken into account.

\paragraph*{Super-resolution optical imaging.}
Quantum estimation theory allows for a rigorous study of fundamental limits in optical microscopy, and serves as a tool for a systematic search for the most precise imaging schemes \cite{Bettens1999,Tsang2016, Lupo2016,Tsang2019,Len2020,Oh2020,deAlmeida2021}. In the elementary scenario, two equally bright incoherent weak point sources are imaged using a translationally invariant system \cite{Tsang2016} (see discussion of limitations of this approximation in \cite{Kurdzialek2022}). The state of a single photon in the image plane is
\begin{equation}
\label{imagingstate}
    \rho_\theta = \frac{1}{2} \left( \ket{u_{+,\theta}}\!\!\bra{u_{+,\theta}} + \ket{u_{-,\theta}}\!\!\bra{u_{-,\theta}} \right),
\end{equation}
where $\braket{x|u_{\pm,\theta}} = u \left( x \pm \theta/2 \right)$, $\{\ket{x}\}$ is the position basis,
$
      \left| u(x) \right|^2 = \left( 2 \pi \sigma^2 \right)^{-1/2} e^{-x^2/2\sigma^2}
$
is the system Point Spread Function. The only unknown parameter is the separation between two sources, $\theta$, the centroid of two points is known \textit{a priori}. Intuitively, it should be hard to estimate $\theta$ when $\theta \ll \sigma$ because then images of two points overlap significantly. This is true for a standard measurement in the position basis $\ket{x}$ because $F_\textrm{C}\left[\{\ket{x}\!\!\bra{x}\} \right] \rightarrow 0$ when $\theta \rightarrow 0$. Surprisingly, the QFI does not depend on $\theta$ at all, $F_\textrm{Q} \left[ \rho_\theta \right] = 1/4\sigma^2$ \cite{Tsang2016}. Therefore, it seems to be no fundamental difference between small and large separations $\theta$, when all quantum measurements are allowed. Unfortunately, this is a highly idealized statement  since the estimation precision for small $\theta$ is highly affected by detection noise, system misalignment, cross-talk noise and other imperfections \cite{Len2020,Oh2020,deAlmeida2021}---in fact, even for the most clever choice of the measurement, the CFI vanishes with $\theta \rightarrow 0$ for all practical scenarios.

At this point, we want to demonstrate the fundamental difficulty of resolving two sources whose images overlap, without referring to any specific noise model, but rather employing the newly introduced FI MeNoS figure of merit. In the most commonly studied super-resolution protocol, the state in the image plane is measured in the basis of orthogonal Hermite-Gaussian modes $\ket{\phi_q}$  whose center lies in the centroid of two observed sources \cite{Tsang2016} ---see \cite{Tsang2019} for an overview of implementations of this measurement based on holography, interferometry, etc. In most of these implementations, we can extract and separate first $K-1$ modes, and the rest of the signal is collected in the $K$-th outcome, such that our POVM consists of elements $M_i=\ket{\phi_i}\!\!\bra{\phi_i}$ for $i \in \{1,...,K-1\}$, $M_K = \mathbb{1} - \sum_{i=1}^{K-1} M_i$.
The CFI increases with $K$, but for $\theta=0$, it approaches the QFI already for $K=2$. The QCRB is saturated in the full range of $\theta$ only for $K \rightarrow \infty$, but the precision is close to optimal for a wide range of $\theta$ already for $K=4$---see Fig.~ \ref{Fig:superresolution}. In the figure we also plot $\chi(\theta)$ for different values of $K$. Unfortunately, $\chi \rightarrow \infty$ for $\theta \rightarrow 0$ in all cases. Consequently, for $\theta \ll \sigma$, it is impossible to achieve the high CFI in the presence of noise using the family of measurements consider so far.

However, there are many other ways to saturate QCRB locally for any fixed value of $\theta$, and some of them may be less susceptible to noise. As for pure states model, we systematically study all QCRB-sat. measurements to find $\chi_\textrm{Q} (\theta)$. Following the technique from \cite{Tsang2016}, we reduce the problem to 4-dimensional Hilbert space $\mathcal{H}^{(\theta)} = \textrm{span} \left\{ \ket{u_{\pm,\theta}}, \partial_\theta \ket{u_{\pm,\theta}} \right\}$, which is a direct sum of two orthogonal 2-dimensional subspaces $\mathcal{H}_\textrm{s}$ and $\mathcal{H}_\textrm{a}$, containing symmetric and antisymmetric modes respectively. We construct orthonormal basis of both subspaces, $ \ket{0}_\textrm{s}, \ket{1}_\textrm{s}$ and $\ket{0}_\textrm{a}, \ket{1}_\textrm{a} $ respectively, such that 
\begin{align}
\label{rhosup}
    &\rho_\theta = \frac{1+\delta}{2} \ket{0}_\textrm{s} \!\! \bra{0} + \frac{1-\delta}{2} \ket{0}_\textrm{a} \!\! \bra{0}, \\
    \label{drhosup}
     &\dot \rho_\theta = \alpha \ket{0}_\textrm{s} \!\! \bra{0} + \beta_\textrm{s} \sigma_x^{(s)} - \alpha \ket{0}_\textrm{a} \!\! \bra{0} + \beta_\textrm{a} \sigma_x^{(a)},
\end{align}
where $\delta = \braket{u_{+,\theta}|u_{-,\theta}}$, real constants $\alpha$, $\beta_\textrm{a}$, $\beta_\textrm{s}$ are specified together with the exact construction of the basis in Appendix \ref{AppE2}. Matrices $\rho_\theta$ and $\dot \rho_\theta$ are both block-diagonal with respect to  $\mathcal{H}_\textrm{s}$ and $\mathcal{H}_\textrm{a}$. This means, that the QCRB-sat. minimal susceptibility POVM contains only elements acting on $ \mathcal{H}_\textrm{s}$ or $\mathcal{H}_\textrm{a}$ (see Appendix \ref{AppD}  for proof). Consider the family of QCRB-sat. POVMs
\begin{equation}
\label{supresoptM}
    \bm{M}_{\varphi_\textrm{s}, \varphi_\textrm{a}} = \left\{ P_{\varphi_\textrm{s}}^{(s)} , P_{\varphi_\textrm{s}+\pi}^{(s)},P_{\varphi_\textrm{a}}^{(a)} , P_{\varphi_\textrm{a}+\pi}^{(a)} \right\},
\end{equation}
where $P_\varphi^{a/s}$ is a projector on $\cos(\varphi/2)\ket{0}_{s/a} + \sin(\varphi/2) \ket{1}_{s/a}$. We prove (Appendix  \ref{AppE2}), that the minimal susceptibility QCRB-sat. measurement is of the the form $\bm{M}_{\varphi_\textrm{s}, \varphi_\textrm{a}}$. Then, we obtain $\chi_\textrm{Q}$ by minimizing numerically $\chi \left[ \bm{M}_{\varphi_\textrm{s}, \varphi_\textrm{a}} \right]$ over $\varphi_\textrm{s}$ and $\varphi_\textrm{a}$, the results are shown in Fig.~\ref{Fig:superresolution} We observe, that $\chi_\textrm{Q} \rightarrow \infty$ when $\theta \rightarrow 0$, which means, that no QCRB-sat. measurement is robust against noise in the region $\theta \ll \sigma$. Surprisingly, $\chi_\textrm{Q}$ does not decrease with $\theta$ everywhere---for example, a minimum $\chi= 4 $ is achieved in $\theta=2\sqrt{2} \sigma$. For $\theta \rightarrow \infty$, $\chi_\textrm{Q} \rightarrow 4$ again, and then the problem is equivalent to a single source localization both from the point of view of $F_\textrm{Q}$ and $\chi_\textrm{Q}$. Interestingly, it is possible to achieve noise susceptibility smaller than $\chi_\textrm{Q}$, when correlations between subsequent photons are present---see Appendix \ref{AppF} for a discussion.

\paragraph*{Outlook.} Computation of the FI MeNoS, should be regarded as a natural sanity-check whenever any idealized quantum metrological protocol is proposed. If this quantity is large (or divergent) this should ring a bell that the performance of the proposed protocol will be significantly reduced by even small imperfection in measurement design. On the contrary, small values indicate that the measurement scheme is robust. The importance of this quantity stems also from the fact, that the FI is a local quantity (computed at single value of parameter) and therefore prone to 
reveal ephemeral effects that vanish in presence of even infinitesimal noise---a property that haunts quantum metrology literature a lot. We envisage that our approach may be naturally extended to multi-parameter estimation framework as well as generalized to cover the Bayesian analysis as well. Still, we expect that in these cases it may be much harder or even impossible to obtain the explicit formula for FI MeNoS analogous to \eqref{chian}.
Moreover, we expect that the role of small measurement disturbances should be less substantial in Bayesian approach, since such an approach is by construction applicable to more realistic scenarios, when the number of collected data samples is finite, and the protocols are expected to perform well beyond the `local estimation approach'.
\begin{acknowledgements}
\paragraph*{Acknowledgements.} We thank Wojciech G{\'o}recki and Janek Ko{\l}ody{\'n}ski for fruitful discussions. This work was supported by the National Science Center (Poland) grant No.\ 2020/37/B/ST2/02134. 
\end{acknowledgements}
\bibliographystyle{apsrev4-2}
\bibliography{bibliography}

\onecolumngrid
\appendix

\section{Noise due to inaccurate meter preparation}
\label{AppA}
In this Section, we prove that an inaccurate initialization of a measuring device M in a state $\rho_\textrm{M}=(1-\epsilon)\ket{0}\!\!\bra{0} + \epsilon \rho_\textrm{M}'$ instead of $\ket{0}\!\!\bra{0}$ changes the POVM acting on S from $\bm{M}$ to $\bm{\tilde M} = (1-\epsilon) \bm{M} + \epsilon \bm{N}$, where the exact form of POVM $\bm{N}$ depends on $\rho_\textrm{M}'$. Initially, S and M are uncorrelated, the state of S is $\rho_\textrm{S}$. The joint state of S and M after unitary interaction $U_\textrm{SM}$ is
\begin{equation}
    \rho_\textrm{SM} = U_\textrm{SM}\rho_\textrm{S} \otimes \rho_\textrm{M} U^\dagger_\textrm{SM} = (1-\epsilon) U_\textrm{SM} \rho_\textrm{S} \otimes \ket{0}\!\!\bra{0} U_\textrm{SM}^\dagger + \epsilon U_\textrm{SM} \rho_\textrm{S} \otimes \rho'_\textrm{M} U_\textrm{SM}^\dagger .
\end{equation}
After interaction, the projective measurement of M in the basis $\ket{i}_{i \in \{1,...,K\}}$ is performed. The probability of obtaining $i$-th outcome is
\begin{equation}
\label{pi01}
    p(i) = \textrm{Tr}\left(  \prescript{}{\textrm{M}}{\braket{i|\rho_\textrm{SM}|i}_\textrm{M}} \right) = (1-\epsilon) p_0(i) + \epsilon p_1(i),
\end{equation}
where

\begin{align}
    &p_0(i) = \textrm{Tr} \left( \rho_\textrm{S} M_i \right),\quad     M_i = \prescript{}{\textrm{M}}{\braket{0|U^\dagger_\textrm{SM}|i}_\textrm{M}\!\!} \braket{i|U_\textrm{SM}|0}_\textrm{M}\\
    &p_1(i) = \textrm{Tr} \left( \rho_\textrm{S} \otimes \mathbb{1} \cdot \mathbb{1} \otimes \rho'_\textrm{M} U_\textrm{SM}^\dagger \ket{i}_\textrm{M}\!\!\bra{i}U_\textrm{SM} \right) = \textrm{Tr} \left( \rho_\textrm{S} \textrm{Tr}_\textrm{M} \left(  \mathbb{1} \otimes \rho'_\textrm{M} U_\textrm{SM}^\dagger \ket{i}_\textrm{M}\!\!\bra{i}U_\textrm{SM} \right) \right) = \textrm{Tr} \left( \rho_\textrm{S} N_i \right). 
\end{align}
The partial scalar product notation was used---$\ket{i}_\textrm{M} \equiv \mathbb{1} \otimes \ket{i}$, $\bra{i}_\textrm{M} \equiv \mathbb{1} \otimes \bra{i}$, partial trace is defined as $\textrm{Tr}_\textrm{M} (\bullet)=\sum_i \prescript{}{\textrm{M}}{\braket{i|\bullet|i}_\textrm{M}}$. Operators $M_i \in \mathcal{L} \left( \mathcal{H}_\textrm{S} \right)$ are the same as those defined in \eqref{eq:prob}, and form undisturbed POVM $\bm{M}$, by $\mathcal{L} \left( \mathcal{H} \right)$ we denote the set of all operators acting on $\mathcal{H}$. Let us now prove that $\bm{N} = \{N_1,...,N_K\}$, where $N_i \in \mathcal{L} \left( \mathcal{H}_\textrm{S} \right) $ is also POVM. Firstly,
\begin{equation}
    \sum_i N_i = \textrm{Tr}_\textrm{M} \left(  \mathbb{1} \otimes \rho'_\textrm{M} U_\textrm{SM}^\dagger \sum_i \ket{i}_\textrm{M}\!\!\bra{i}U_\textrm{SM} \right) = \textrm{Tr}_\textrm{M} \left( \mathbb{1} \otimes \rho'_\textrm{M}\right) = \mathbb{1}_\textrm{S}.
\end{equation}
Moreover,
\begin{equation}
    N_i = \textrm{Tr}_\textrm{M} \left( K_i^\dagger K_i \right),\quad     K_i=\prescript{}{\textrm{M}}{\bra{i}}~ U_\textrm{SM}~ \mathbb{1}\otimes \sqrt{\rho'_\textrm{M}},
\end{equation}
so $N_i \ge 0$ because partial trace of positive operator is positive. From \eqref{pi01},
\begin{equation}
    p(i) = (1-\epsilon) \textrm{Tr}(\rho_\textrm{S} M_i) + \epsilon \textrm{Tr} (\rho_\textrm{S} N_i) = \textrm{Tr}(\rho_\textrm{S} \tilde M_i),
\end{equation}
so indeed $\bm{M}$ is replaced with $\bm{ \tilde M}$ due to the considered disturbance of the initial state of a measuring device.

\section{Different sources of noise---examples}
\label{AppB}
In practice, the real source of noise may be different than the one described in the previous section. However, our general noise model $\bm{M} \rightarrow (1-\epsilon) \bm{M} + \epsilon \bm{N}$ encompasses many different types of noise. In this Section, we demonstrate a few practical examples of noise types together with corresponding noise POVMs $\bm{N}$.
\subsection{Signal losses}
Let us consider arbitrary POVM with $K-1$ elements, $\bm{M'} = (M_2,...,M_K)$---clearly, the measurement described by $\bm{M'}$ is equivalent to the one described by $\bm{M} = (0,M_2,...,M_K) $ because the probability associated with the 1st element of $\bm{M}$ is always $0$. When the noise of the form $\bm{N} = (\mathbb{1},0,...,0)$ acts with probability $\epsilon$, then the resulting POVM is $\bm{\tilde M} = \left( \epsilon \mathbb{1}, (1-\epsilon)M_2,..., (1-\epsilon) M_K \right)$. The 1st element of $\bm{\tilde M}$ corresponds to non-informative outcome because its probability doesn't depend on $\rho$ and is always equal to $\epsilon$. Therefore, $\bm{\tilde M}$ describes the scenario in which we loose our signal with probability $\epsilon$, independently of the measured state. Then, $G[\bm{N}]=0$, and consequently $\chi[\bm{M}, \bm{N}]=1$---for example, when $\epsilon=1 \%$, then we loose $1 \%$ of the signal, which leads to $\epsilon \chi [\bm{M}, \bm{N} ] = 1 \%$ relative decrease of the Fisher Information.
\subsection{Dark counts}
Let us assume, that we have $K$ detectors in our setup, undisturbed measurement is described by POVM $\bm{M} = (M_1,...,M_K)$. Detector may sometimes ''click'' even if no signal was sent---such events are called ''dark counts'', in each repetition real signal is observed with probability $1- \epsilon$, and a dark count with probability $\epsilon$. Probabilities of dark counts do not depend on $\rho$, but may be different for different detectors, so the most general POVM describing dark counts is $\bm{N} = \left(q_1 \mathbb{1}, ..., q_K \mathbb{1} \right)$, $q_1+...+q_K=1$. Dark counts are generally more problematic than signal losses because noise is not separated from the signal. This is reflected by larger $\chi[\bm{M}, \bm{N}]$---notice that $G[\bm{N}] = \sum_i q_i l_i^2>0$, where $l_i = \frac{\partial_\theta p_i}{p_i}$, $p_i = \textrm{Tr}(\rho_\theta M_i)$. For a particular case $q_i = p_i$, we have $G[\bm{N}] = F_\textrm{C}[\bm{M}]$, so $\chi[\bm{M}, \bm{N}] = 2$---there are dark counts models, for which the decrease of FI is twice as large as for signal losses. 
\subsection{Cross-talks}
This type of noise appears, when different measurement outcomes are sometimes confused with each other. In a noiseless scenario, $i$-th detector clicks with a probability $p_i = \textrm{Tr}(\rho_\theta M_i)$ ($i \in \{1,...,K\}$). However, because of cross-talks, the result associated with $M_j$ is interpreted as $i$ with a probability $t(i|j)$, so the probability of obtaining $i$-th outcome becomes $\tilde p_i = \sum_{j=1}^K t(i|j) p_j = \textrm{Tr} (\rho_\theta \tilde M_i)$, where operators $\tilde M_i = \sum_j t(i|j) M_j$ form a noise-affected POVM $\bm{\tilde M}$,  $\sum_i t(i|j) = 1$ for all $j$.  As in all previous examples, we assume that the noise is $\epsilon$-small, so $t(i|i) \ge 1-\epsilon$ for all $i$. Then we can express cross-talk probabilities as $t(i|j) = (1- \epsilon) \delta_{ij} + \epsilon \tilde t(i|j)$, where $\tilde t (i|j)>0$, $\sum_i \tilde t(i|j) = 1$. Consequently, $ \tilde M_i = (1-\epsilon) M_i + \epsilon N_i$, where operators $N_i = \sum_j \tilde t(i|j) M_j $ form a noise POVM $\bm{N}$.

\subsection{Random rotations of measurement basis}
Let us consider a projective measurement $\bm{M} = \left( \ket{\phi_1}\!\!\bra{\phi_1},..., \ket{\phi_K}\!\! \bra{\phi_K} \right)$. In some cases, the misalignment of the measuring device may change the measurement basis. Let us assume, that the basis is rotated by unitary operation $\hat U_j$ with probability $\epsilon q_j$, $\sum_j q_j = 1$. Then, the noise-affected POVM is $\bm{\tilde M} = (1-\epsilon) \bm{M} + \epsilon \bm{N}$, where $\bm{N}$ consists of elements $N_i = \sum_j q_j U_j \ket{\phi_i}\!\!\bra{\phi_i} U_j^\dagger$.
\section{Explicit Formula for FI MeNoS}
\label{AppC}
In this Section, the derivation of \eqref{chian} is completed. As we proved in the main text, the worst-case scenario noise is of the form $\bm{\tilde N} = (\tilde N_1,0,...,0,\mathbb{1}-\tilde N_1)$, and the maximization of $G[\bm{N}]$ over all POVMs $\bm{N}$ simplifies to
\begin{equation}
  \label{maxsimp}
      \underset{\bm{N} \in \mathcal{M}}{\textrm{max}} G[\bm{N}] = \underset{0 \le \tilde N_1 \le \mathbb{1}}{\textrm{max}} \textrm{Tr}\left[A_1 \tilde N_1 + A_K \left(\mathbb{1}-\tilde N_1 \right) \right]=l_K^2 + \underset{0 \le \tilde N_1 \le \mathbb{1}}{\textrm{max}} \textrm{Tr}\left[ (A_1-A_K) \tilde N_1 \right].
\end{equation}

Let $\ket{e_i}$ be orthonormal basis diagonalizing $A_1-A_K$. We can express both $A_1-A_K$ and $\tilde N_1$ in this basis:
\begin{equation}
    A_1 - A_K = \sum_{i=1}^R \lambda_i \ket{e_i}\!\!\bra{e_i},~~ \tilde N_1 = \sum_{i,j=1}^R \tilde N_{ij} \ket{i}\!\!\bra{j},
\end{equation}
where $R=\textrm{dim}\mathcal{H}_\textrm{S}$ is the size of the matrices. We choose the ordering of $\ket{e_i}$ such that for some $P \in \{1,...,R\}$, $\lambda_1,....,\lambda_P \ge 0$ and $\lambda_{P+1},...,\lambda_R <0$. The maximization problem from the RHS of \eqref{maxsimp} transforms to
\begin{equation}
    \underset{0 \le \tilde N_1 \le \mathbb{1}}{\textrm{max}} \textrm{Tr}\left[ (A_1-A_K) \tilde N_1 \right] = \underset{\tilde N_{ii}}{\textrm{max}} \sum_i \tilde N_{ii} \lambda_i.
\end{equation}
From positivity of $\tilde N_1$ and of $\mathbb{1}- \tilde N_1$, $\tilde N_{ii} \ge 0$ and $\tilde N_{ii} \le 1$, therefore
\begin{equation}
    \underset{0 \le \tilde N_1 \le \mathbb{1}}{\textrm{max}} \textrm{Tr}\left[ (A_1-A_K) \tilde N_1 \right] = \sum_{i=1}^P \lambda_i,
\end{equation}
which is the sum of positive eigenvalues of $A_1-A_K$. Moreover, we have
$
    \textrm{Tr} (A_1 - A_K) = \sum_{i=1}^P \lambda_i + \sum_{i=P+1}^R \lambda_i$,  $\|A_1-A_K\|_1 = \sum_{i=1}^P \lambda_i - \sum_{i=P+1}^R \lambda_i$,
which means that
\begin{equation}
     \sum_{i=1}^P \lambda_i = \frac{1}{2} \left(\textrm{Tr}(A_1-A_K) + \|A_1-A_K\|_1 \right) = \frac{1}{2}(l_1^2-l_K^2+ \|A_1-A_K\|_1).
\end{equation}
Taking all this results together, we obtain
\begin{equation}
\label{maxG}
   \underset{\bm{N} \in \mathcal{M}}{\textrm{max}} G[\bm{N}] = \underset{0 \le \tilde N_1 \le \mathbb{1}}{\textrm{max}} \textrm{Tr}\left[A_1 \tilde N_1 + A_K \left(\mathbb{1}-\tilde N_1 \right) \right]  = \frac{1}{2}(l_1^2+l_K^2+ \|A_1-A_K\|_1),
\end{equation}
which, after inserting into \eqref{chiMN2}, leads to \eqref{chian}.
\section{The most robust among the most informative measurements---Theorems}
\label{AppD}
In this Section, we show how to approach the problem of finding a QCRB-sat. measurement with the minimal FI MeNoS. Before showing solutions for examples mentioned in the main text, let us introduce some theorems that help to systematically describe potential candidates for such a measurement. Let us start with quoting a condition for QCRB-saturability, firstly introduced in \cite{Braunstein1994}, and then reformulated in \cite{Zhou2020} in a following form:

\vspace{0.5\baselineskip}
\noindent
\textbf{Theorem 1.}
POVM $\bm{M} = \{M_i\}_i$ is QCRB-sat. for the family of states $\rho_\theta$ around $\theta=\theta_0$ iff 
\begin{equation}
\label{sat1}
     M_i^{1/2} L_{jk} M_i^{1/2} = 0,~ \forall i,j,k
\end{equation}
and
\begin{equation}
\label{sat2}
    \forall i~ \textrm{s.t.}~ \textrm{Tr}\left( M_i \rho_{\theta_0} \right)=0,~~ M_i^{1/2} \Lambda_{\theta_0} \ket{\psi_{{\theta_0},j}}=0  ,~\forall j.
\end{equation}
Here $\rho_\theta = \sum_j p_{\theta,j} \ket{\psi_{\theta,j}} \!\! \bra{\psi_{\theta,j}}$ is the diagonalization of $\rho_\theta$, $p_{\theta,j} > 0$, $\Lambda_\theta$ is the symmetric logarithmic derivative matrix, and $L_{jk} =  \ket{\psi_{\theta_0,j}} \!\! \bra{\psi_{\theta_0,k}} \Lambda_{\theta_0}- \Lambda_{\theta_0} \ket{\psi_{\theta_0,j}} \!\! \bra{\psi_{\theta_0,k}} $.

\vspace{0.5\baselineskip}
\noindent
\emph{Proof.} See Ref.\cite{Zhou2020} for a proof. $\blacksquare$
\vspace{0.5\baselineskip}

Next two lemmas will help us to reduce the set of potential candidates for the minimal MeNoS QCRB-sat. measurements to those containing only rank-one elements $M_i$ (see Theorem 2). 

\vspace{0.5\baselineskip}
\noindent
\textbf{Lemma 1} Let $p_\theta(i)$ be the family of probability distributions indexed by $\theta$, $i \in \{1,...,K\}$. Let $ q_\theta(j) = \sum_i p_\theta(i) t(j|i)$ be another probability distribution, $j \in \{1,...,L\}$, $t(j|i) \ge 0$, $\sum_j t(j|i)=1$. Then
\begin{align}
        1.&~F_\textrm{C}[p_\theta(i)] \ge F_\textrm{C}[q_\theta(j)], \\
        2.&~F_\textrm{C}[p_\theta(i)] = F_\textrm{C}[q_\theta(j)] ~ \textrm{iff}  ~\forall_{j,i_1,i_2} \left( t(j|i_1) \ne 0 \land t(j|i_2) \ne 0 \right) \Rightarrow l_{i_1} = l_{i_2},  
    \end{align}
where $l_i = \frac{ \dot p_\theta(i)}{p_\theta(i)}$. Point 2. means, that random mixing of results does not decrease the CFI iff only results with the same logarithmic derivatives are mixed with each other.

\vspace{0.5\baselineskip}
\noindent
\emph{Proof.}
Let us adopt a short-hand notation $p_i \equiv p_\theta(i)$, $q_j \equiv q_\theta(j)$. The CFI associated with the distribution $q_\theta(j)$ can be written as
\begin{equation}
\label{FcFj}
    F_\textrm{C} \left[q_\theta(j) \right] = \sum_{j=1}^L F_j,~\textrm{where}~ F_j = \frac{\dot q_j^2}{q_j}.
\end{equation}

Now we want to prove, that
\begin{equation}
\label{Fjin}
    F_j = \frac{\left(\sum_{i=1}^K t(j|i) \dot p_i \right)^2}{q_j} \le \sum_{i=1}^K t(j|i) \frac{\dot p_i^2}{p_i}.
\end{equation}
For a fixed $j$, let us introduce the notation $\alpha_i = \frac{t(j|i) p_i}{q_j}$. From the definition of $q_j$ follows that $\sum_i \alpha_i =1$, moreover $\alpha_i \ge 0$. Taking this into account, we have
\begin{equation}
\label{ineq12}
   \left|\sum_{i=1}^K t(j|i) \dot p_i \right| = \left| \sum_{i=1}^K \alpha_i\frac{t(j|i) \dot p_i}{ \alpha_i} \right| \overset{(1)}{\le} \sum_{i=1}^K \alpha_i \left|\frac{t(j|i) \dot p_i}{ \alpha_i} \right| \overset{(2)}{\le}  \sqrt{\sum_{i=1}^K \alpha_i \left( \frac{t(j|i) \dot p_i}{ \alpha_i} \right)^2},
\end{equation}
where (1) is triangle inequality, and (2) follows from the weighted power mean inequality (between arithmetic and quadratic mean, AM-QM in short).
After squaring both sides of \ref{ineq12}, substituting the definition of $\alpha_i$ in the RHS, and dividing both sides by $q_j$ we obtain inequality \ref{Fjin}. Subsequently, using \ref{Fjin} and \ref{FcFj} , we obtain
\begin{equation} \label{FciFcj}
    F_\textrm{C} \left[q_\theta(j) \right] \le \sum_{j=1}^L \sum_{i=1}^K t(j|i) \frac{\dot p_i^2}{p_i} = \sum_{i=1}^K \frac{\dot p_i^2}{p_i} = F_\textrm{C} \left[p_\theta(i)\right],
\end{equation} 
which is exactly part 1. of our Lemma. In order to prove part 2., let us notice, that inequality \ref{FciFcj} is saturated iff inequalities (1) and (2) in \ref{ineq12} are saturated for all $j$. For a given $j$, part (1) of \ref{ineq12} becomes equality iff for all $i$ s.t. $t(j|i) \ne 0$, terms $\frac{t(j|i) \dot p_i}{ \alpha_i}= q_j \frac{\dot p_i}{p_i}$ have the same sign. Part (2), which is AM-QM inequality becomes saturated, when all terms in the mean with non-zero weights are equal, which implies that for a given $j$, for all all $i$ s.t. $t(j|i) \ne 0$ terms $\left|q_j \frac{\dot p_i}{p_i}\right|$ are equal. From these two saturability conditions, we obtain part (2) of Lemma 1. $\blacksquare$
\vspace{0.5\baselineskip}

The next lemma states that random mixing of the elements of POVM either decreases the CFI or keeps both the CFI and FI MeNoS unaffected.

\vspace{0.5\baselineskip}
\noindent
\textbf{Lemma 2.}
Let $\bm{M}, \bm{M}' \in \mathcal{M}$, $\bm{M}=\{M_i\}_{i \in \{1,...,K\}}$, $\bm{M}'=\{M_j'\}_{j \in \{1,...,L\}}$, $M_j' = \sum_i t(j|i) M_i$, where $t(j|i) \ge 0$, $\sum_j t(j|i) = 1$. The family of quantum states $\rho_\theta$ is fixed. Then
\begin{align}
        1.&~F_\textrm{C}[\bm{M}] \ge F_\textrm{C}[\bm{M}'], \\
        2.&~F_\textrm{C}[\bm{M}] = F_\textrm{C}[\bm{M}'] \Rightarrow \chi \left[ \bm{M} \right] = \chi \left[ \bm{M}' \right].  
    \end{align}
    
\noindent    
\emph{Proof.}
The classical probability distributions associated with measurements $\bm{M}$ and $\bm{M}'$ are $p_\theta(i)= \textrm{Tr} (\rho_\theta M_i)$, $q_\theta(j) = \textrm{Tr} (\rho_\theta M_j) $. From the definition of $\bm{M}'$, we have $q_\theta(j) = \sum_i t(j|i) p_\theta(i)$. Therefore, part 1. of Lemma 2 is a direct consequence of part 1. of Lemma 1.

The logarithmic derivative of $q_j$ is
\begin{equation}
    l'_j = \frac{\dot q_j}{q_j} = \frac{\sum_i t(j|i) \dot p_i}{\sum_i t(j|i) p_i} = \frac{\sum_i t(j|i) l_i p_i}{\sum_i t(j|i) p_i}
\end{equation}
When $F_\textrm{C} [\bm{M}] = F_\textrm{C} [\bm{M}'] $, then, according to part 2. of Lemma 1.,
\begin{equation}
     l_{j_1}=l_{j_2}=...=l_{j_{K_j}}=l_j',
\end{equation}
where $j_1,j_2,...j_{K_j}$ are the indices satisfying $t(j|j_i) \ne 0$. That means, that the sequence of logarithmic derivatives $l'_j$ consist off exactly the same elements as the sequence $l_i$ (some elements may repeat), and the minimal and maximal logarithmic derivatives for $\bm{M}$ and $\bm{M}'$ are the same, so according to \eqref{chian}, $\chi \left[ \bm{M} \right] = \chi \left[ \bm{M}' \right]$. $\blacksquare$

\vspace{0.5\baselineskip}
\noindent
\textbf{Theorem 2.}
For any family of quantum states $\rho_\theta$, there exists a QCRB-sat. measurement with a minimal FI MeNoS whose all elements are rank-one matrices.

\vspace{0.5\baselineskip}
\noindent
\emph{Proof.} Let $\bm{M}' = \{M'_i\}_i $ be any QCRB-sat. POVM with a minimal FI MeNoS, $F_\textrm{C} \left[ \bm{M}' \right] = F_\textrm{Q}$, $\chi [\bm{M}'] = \chi_\textrm{Q}$. In a diagonal form, $M'_i = \sum_j \lambda_{i,j} \ket{\phi_{i,j}}\!\!\bra{\phi_{i,j}}$, $\lambda_{i,j} \ge 0$ because $M'_i$ are positive-semidefinite. Let us consider a POVM $\bm{M} = \{\lambda_{i,j}\ket{\phi_{i,j}}\!\!\bra{\phi_{i,j}}\}_{i,j}$, whose elements are all rank-one matrices. According to Lemma 2, $F_\textrm{C} \left[\bm{M}\right] \ge F_\textrm{C} \left[\bm{M}'\right] = F_\textrm{Q} $, but on the other hand, the CFI cannot be greater then the QFI, so $F_\textrm{C} \left[\bm{M}\right]=F_\textrm{C} \left[\bm{M}'\right]= F_\textrm{Q} $. Therefore, from part 2. of Lemma 2., $\chi[\bm{M}] = \chi[\bm{M}']= \chi_\textrm{Q}$, which means, that $\bm{M}$ is a QCRB-sat. POVM with a minimal MeNoS consisting of rank-one matrices only. $\blacksquare$
\vspace{0.5\baselineskip}

The next theorem is helpful for problems involving states with block-diagonal structure, it will be used in the next section to study super-resolution imaging.

\vspace{0.5\baselineskip}
\noindent
\textbf{Theorem 3.}
Let $\rho_\theta, \dot \rho_\theta \in \mathcal{L}(\mathcal{H})$ be a density matrix and its derivative, which can be decomposed as
\begin{align}
    &\rho_{\theta} = \rho_{\theta}^{(1)} + \rho_{\theta}^{(2)} + ... + \rho_{\theta}^{(L)}, \\
    & \dot \rho_{\theta} = \dot \rho_{\theta}^{(1)} + \dot \rho_{\theta}^{(2)} + ... + \dot \rho_{\theta}^{(L)}, 
\end{align}
where $\rho_{\theta}^{(l)}, \dot \rho_{\theta}^{(l)}$ act on a subspace $\mathcal{H}^{(l)} \subset \mathcal{H}$ (it has zeros outside this subspace), $\mathcal{H} = \mathcal{H}^{(1)} \oplus \mathcal{H}^{(2)} \oplus ... \oplus \mathcal{H}^{(L)}$, subspaces $\mathcal{H}^{(l)}$ are orthogonal to each other. Then, there exists a QCRB-sat. measurement with a minimal MeNoS, which can be written as \begin{equation}
\label{POVMsep}
\bm{M} = \left\{M_1^{(1)},...,M_{K_1}^{(1)},M_1^{(2)},...,M_{K_2}^{(2)},...,M_1^{(L)},...,M_{K_L}^{(L)}, \right\} \end{equation}
 where for each $l \in \{1,...,L\}$ matrices $M_1^{(l)},...,M_{K_l}^{(l)}$ act on   $\mathcal{H}^{(l)} $ and form a QCRB-sat. POVM for a family of normalized states $\tilde \rho_{\theta}^{(l)} = \rho_{\theta}^{(l)}/\textrm{Tr} \left( \rho_{\theta}^{(l)} \right)$.

\vspace{0.5\baselineskip}
\noindent
\emph{Proof.}
Let us denote an orthogonal projector on $\mathcal{H}^{(l)}$ by $P^{(l)}$. Let $\bm{M}' = \{M'_i\}_{i \in \{1,...,K\}} $ be any QCRB-sat. POVM with a minimal FI MeNoS. Let us construct another POVM $\bm{M}'' = \{M''_i\}_{i \in \{1,...,K\}}$, where 
$
     M''_i = P^{(1)} M_i' P^{(1)}+P^{(2)} M_i' P^{(2)}+...+P^{(L)} M_i' P^{(L)}. 
$

We have
\begin{equation}
    \textrm{Tr}(\rho_{\theta} M_i') = \textrm{Tr}\left(\left(\rho_{\theta}^{(1)} + \rho_{\theta}^{(2)}+...+\rho_{\theta}^{(L)} \right) M_i' \right) = \textrm{Tr} \left( \rho_{\theta} \left( P^{(1)} M_i' P^{(1)} + ... +P^{(L)} M_i' P^{(L)} \right) \right) = \textrm{Tr} \left( \rho_{\theta} M_i'' \right), 
\end{equation}
where we used the fact that $\rho_{\theta}^{(l)} = P^{(l)} \rho_{\theta} P^{(l)} $.
Similarly,
\begin{equation}
     \textrm{Tr}(\dot \rho_{\theta} M_i') = \textrm{Tr}\left(\left(\dot \rho_{\theta}^{(1)} + \dot \rho_{\theta}^{(2)}+...+\dot \rho_{\theta}^{(L)} \right) M_i' \right) =\textrm{Tr} \left( \dot \rho_{\theta} M_i'' \right).
\end{equation}
Therefore, POVMs $\bm{M}'$ and $\bm{M}''$ are fully equivalent---probabilities of different outcomes $p_{\theta}(i)$ and their derivatives $\dot p_{\theta}(i)$ are the same for $\bm{M}'$ and $\bm{M}''$. Consequently, $F_\textrm{C} [ \bm{M}']=F_\textrm{C} [ \bm{M}'']$ and $\chi [ \bm{M}']=\chi [ \bm{M}'']$. 

Let us finally construct a POVM
\begin{equation}
    \bm{M} = \left\{ P^{(1)} M_1' P^{(1)}, P^{(1)} M_2' P^{(1)},...,P^{(1)} M_K' P^{(1)},..., P^{(L)} M_1' P^{(L)},P^{(L)} M_2' P^{(L)},...,P^{(L)} M_K' P^{(L)}\right\},
\end{equation}
which is clearly of the form from \eqref{POVMsep} because $P^{(l)} M_i' P^{(l)}$ act on $  \mathcal{H}^{(l)}$. Moreover, elements of $\bm{M}''$ are linear combinations of the elements of $\bm{M}$, so, according to Lemma 2, $F_\textrm{C} \left[ \bm{M} \right] = F_\textrm{Q}$ and $\chi [\bm{M}] = \chi_\textrm{Q}$. Therefore, we constructed a QCRB-sat. POVM with a minimal MeNoS which has a form defined in \eqref{POVMsep}. Let us now prove, that when $\bm{M}$ from \eqref{POVMsep} is QCRB-sat., then for any $l \in \{1,...,L\}$, POVM $\bm{M}^{(l)} = \{ P^{(l)} M_1' P^{(l)}, P^{(l)} M_2' P^{(l)},...,P^{(l)} M_K' P^{(l)}  \}$ is QCRB-sat. for $\tilde \rho_{\theta}^{(l)}$. It is straightforward to show that 
\begin{equation}
\label{Fcmany}
    F_\textrm{C} \left[ \rho_\theta, \bm{M} \right] = \sum_{l=1}^L F_\textrm{C} \left[  \tilde \rho_\theta^{(l)}, \bm{M}^{(l)} \right] + F_\textrm{C} \left[ P_\theta(l) \right],
\end{equation}
where $P_\theta(l) = \textrm{Tr} \left( \rho_\theta^{(l)} \right)$ is the probability of measuring $\rho_\theta$ in a subspace $\mathcal{H}^{(l)}$. 
When for some $l$, $\bm{M}^{(l)}$ is not QCRB-sat., then there exists a POVM $\bm{M}_2^{(l)}$ for which $F_\textrm{C} \left[  \tilde \rho_\theta^{(l)}, \bm{M}_2^{(l)} \right] > F_\textrm{C} \left[  \tilde \rho_\theta^{(l)}, \bm{M}^{(l)} \right]$. Then, according to \eqref{Fcmany}, $F_\textrm{C} \left[ \rho_\theta, \bm{M}_2 \right]>F_\textrm{C} \left[ \rho_\theta, \bm{M} \right]$, where $\bm{M}_2 = \left\{ \bm{M}^{(1)},...,\bm{M}_2^{(l)},...,\bm{M}^{(K)} \right\}$. This is a contradiction with the assumption that $\bm{M}$ is QCRB-sat.---therefore $\bm{M}^{(l)}$ is QCRB-sat. for $\tilde \rho_\theta^{(l)}$. $\blacksquare$

\vspace{0.5 \baselineskip}
\noindent
\textbf{Theorem 4.}
Function $\chi[\bm{M}]$ defined in \eqref{chian} is  non-decreasing with $l_K$ and  non-increasing with $l_1$ for fixed $F_\textrm{C}[\bm{M}]$ (when $l_1$, $l_K$, $F_\textrm{C}[\bm{M}]$ are treated as independent variables).

\vspace{0.5 \baselineskip}
\noindent
\emph{Proof.} 
Let us introduce a notation
\begin{equation}
    G[\bm{N},l_1,...,l_K] = \sum_{i=1}^K \textrm{Tr}(A_i N_i),~~A_i=l_i^2 \rho_\theta - 2 l_i \dot \rho_\theta,
\end{equation}
which is a more precise version of \eqref{Gdef}. We assume that $\rho_\theta$ and $\dot \rho_\theta$ are fixed, so we do not write them down explicitly as arguments of $G$. Let  $l_1 \le ...\le l_K \le l_{K+1} $ be an arbitrary ascending sequence of real numbers. For any $K$-element POVM $\bm{N}=\{N_1,...,N_K\}$ it is possible to construct a $K+1$-element POVM $\bm{\tilde N}$ satisfying
\begin{equation}
    G\left[ \bm{N},l_1,...,l_K \right] = G \left[ \bm{\tilde N},l_1,...,l_K,l_{K+1} \right],
\end{equation}
just by defining $\bm{\tilde N} = \{N_1,...,N_K,0\}$. Consequently,
\begin{equation}
    \underset{\bm{N} \in \mathcal{M}}{\textrm{max}} G\left[ \bm{N},l_1,...,l_K \right] \le \underset{\bm{\tilde N} \in \mathcal{M}}{\textrm{max}} G\left[ \bm{\tilde N},l_1,...,l_K,l_{K+1} \right].
\end{equation}
After applying \eqref{maxG} to both sides of this inequality, we obtain
\begin{equation}
\frac{1}{2}(l_1^2+l_K^2+ \|A_1-A_K\|_1) \le \frac{1}{2}(l_1^2+l_{K+1}^2+ \|A_1-A_{K+1}\|_1),
\end{equation}
which means that for a fixed $F_\textrm{C} [\bm{M}]$, $\chi[\bm{M}]$ is non-decreasing with $l_{K+1}$ because real number $l_K \le l_{K+1}$ can be arbitrary. Analogously, we can prove that for any pair $l_0 \le l_1$,
\begin{equation}
    \frac{1}{2}(l_1^2+l_K^2+ \|A_1-A_K\|_1) \le \frac{1}{2}(l_0^2+l_{K}^2+ \|A_0-A_{K}\|_1),
\end{equation}
from which follows the 2nd part of our thesis. $\blacksquare$

\section{The most robust among the most informative measurements---Examples}
\label{AppE}
\subsection{Pure state models}
\label{AppE1}
Let $\rho_\theta = \ket{\psi_\theta} \!\! \bra{\psi_\theta}$, $\dot \rho_\theta = \ket{\psi_\theta}\!\! \bra{\dot \psi_\theta} + \ket{\dot \psi_\theta}\!\! \bra{\psi_\theta}$. Then, the symmetric logarithmic derivative matrix is $\Lambda_\theta = 2 \left( \ket{\psi_\theta}\!\! \bra{\dot \psi_\theta} + \ket{ \dot \psi_\theta}\!\! \bra{ \psi_\theta} \right) $, and the QFI is
\begin{equation}
    F_\textrm{Q} = \textrm{Tr} (\rho_\theta \Lambda_\theta^2) = 4 \left( \braket{\dot \psi_\theta | \dot \psi_\theta} - \left| \braket{\dot \psi_\theta | \psi_\theta} \right|^2  \right).
\end{equation}
For a fixed value of a parameter $\theta=\theta_0$ we define 
\begin{align}
\label{canb1}
    &\ket{0} = \frac{1}{\sqrt{2}} \ket{\psi_{\theta_0}} + i \sqrt{2 F_\textrm{Q}^{-1}} \left( \ket{ \dot \psi_{\theta_0}} - \braket{\psi_{\theta_0}|\dot \psi_{\theta_0}} \ket{\psi_{\theta_0}} \right), \\
    \label{canb2}
    &\ket{1} =\frac{1}{\sqrt{2}} \ket{\psi_{\theta_0}} - i \sqrt{2 F_\textrm{Q}^{-1}} \left( \ket{ \dot \psi_{\theta_0}} - \braket{\psi_{\theta_0}|\dot \psi_{\theta_0}} \ket{\psi_{\theta_0}} \right).
\end{align}
It is straightforward to check that $\ket{0}$ and $\ket{1}$ form o.-n. basis of $\textrm{span} \left\{ \ket{\psi_{\theta_0}}, \ket{\dot \psi_{\theta_0}} \right\}$,  the representations of $\rho_{\theta_0}$ , $\dot \rho_{\theta_0}$ and $\Lambda_{\theta_0}$ in this basis are:
\begin{equation}
\label{rhocan}
    \rho_{\theta_0} = \frac{1}{2} \begin{bmatrix} 1 & 1 \\ 1 & 1 \end{bmatrix} = \ket{+}\!\! \bra{+}, ~~
    \dot \rho_{\theta_0} = \frac{1}{2} \sqrt{F_\textrm{Q}} \begin{bmatrix} 0 & -i \\ i & 0 \end{bmatrix} = \frac{1}{2} \sqrt{F_\textrm{Q}} \sigma_y, ~~ \Lambda_{\theta_0} = \sqrt{F_\textrm{Q}} \sigma_y.
\end{equation}
The condition \eqref{sat1} from Theorem 1 says, that all elements $M_i$ of any QCRB-sat. POVM must satisfy 
\begin{equation}
\label{condQCRB}
    M_i^{1/2} L_{11} M_i^{1/2} = 0,
\end{equation}
where 
\begin{equation}
    L_{11} = \ket{\psi_{\theta_0}}\!\! \bra{\psi_{\theta_0}} \Lambda_{\theta_0} - \Lambda_{\theta_0} \ket{\psi_{\theta_0}}\!\! \bra{\psi_{\theta_0}}   = i \sqrt{F_\textrm{Q}} \sigma_z
\end{equation}
in the basis $\ket{0},\ket{1}$.
After taking the trace of $n$-th power of both sides of \ref{condQCRB}, we obtain
\begin{equation}
\label{condQCRB2}
    \textrm{Tr} \left( (M_i L_{11})^n \right)=0.
\end{equation}
The most general parametrization of $M_i$ is
\begin{equation}
    M_i = \begin{bmatrix} \alpha_i & \beta_i - i \gamma_i \\ \beta_i + i \gamma_i & \delta_i \end{bmatrix},
\end{equation}
where $\alpha_i, \beta_i, \gamma_i, \delta_i \in \mathbb{R}$. Using \ref{condQCRB2} with $n=1$ and $n=2$, we obtain the following conditions:
\begin{equation}
    n=1: \alpha_i = \delta_i,~~
    n=2:  \det M_i = 0.
\end{equation}
From the 2nd condition, one eigenvalue of $M_i$ must be $0$, so $M_i$ is rank-one, and can be written as
\begin{equation}
\label{POVMproj}
    M_i = \lambda_i \ket{\phi_i}\!\! \bra{\phi_i},~~
    \ket{\phi_i} = \begin{bmatrix} a_i \\ b_i \end{bmatrix} = \frac{1}{\sqrt{2}} \begin{bmatrix} 1 \\ e^{i \varphi_i} \end{bmatrix} ,
\end{equation}
 $0 \le \lambda_i \le 1$, $\left|a_i\right| = \left|b_i\right|=1/\sqrt{2}$ because $\alpha_i =\delta_i$ and because of normalization of $\ket{\phi_i}$. We obtained necessary conditions for QCRB saturation, but it turns out, that they are also sufficient---all POVMs, whose elements are of the described form, saturate QCRB. To show this, let us compute the CFI explicitly:
\begin{equation}
    F_\textrm{C} = \sum_i \lambda_i \frac{\left(\braket{\phi_i|\dot \rho_{\theta_0} |\phi_i} \right)^2}{\braket{\phi_i| \rho_{\theta_0} |\phi_i}} = \frac{F_\textrm{Q}}{2} \sum_i \lambda_i \left( 1 -a_i \bar b_i - \bar a_i b_i \right) = F_\textrm{Q},
\end{equation}
where we used the identity $\sum_i M_i = \mathbb{1}$, from which follows that $\sum_i \lambda_i = \textrm{Tr} (\mathbb{1}) = 2$, and $\sum_i \lambda_i \bar a_i b_i = \sum_i \lambda_i a_i \bar b_i =0$. To conclude, for a given parametrization of a family of pure states, a POVM is QCRB-sat. iff it contains only elements proportional to projectors on states from the Bloch sphere's equator.

At this point, we are ready to compare different QCRB-sat. measurements according to their FI MeNoS. Let us start with projective measurements, i.e. those for which $K=2$, $\lambda_1 = \lambda_2 =1$, $\varphi_1 = -\varphi$, $\varphi_2 = \pi - \varphi$, we refer to notation from \eqref{POVMproj}, $\varphi \in [0, 2 \pi ]$ corresponds to phase $\varphi$ which has to be added to the lower arm of Mach-Zender interferometer from Fig.~\ref{Fig:interferometer} in order to implement the described projective measurement. Logarithmic derivatives associated with this measurement are
\begin{equation}
    l_1 = \frac{\textrm{Tr} (\dot \rho _{\theta_0} M_1)}{\textrm{Tr} ( \rho _{\theta_0} M_1)} = \sqrt{F_\textrm{Q}} \tan (\varphi/2),~~l_2 = \frac{\textrm{Tr} (\dot \rho _{\theta_0} M_2)}{\textrm{Tr} ( \rho _{\theta_0} M_2)} = -\sqrt{F_\textrm{Q}} \tan^{-1}(\varphi/2). 
\end{equation}
There are only two logarithmic derivatives, so one of them is maximal, and another is minimal. After inserting the values of $l_1$, $l_2$ into $\eqref{chian}$ and computing eigenavalues of  $A_1-A_2$ we obtain \eqref{chiint}, from which it follows that $\chi[\bm{M}] \ge 4$ for all QCRB-sat. projective measurements, and the inequality is saturated only for $\varphi = \pi/2$. Let us now show, that $\chi[\bm{M}] \ge 4$ for all QCRB-sat. measurements (not only projective). For $\bm{M}$ with $K$ elements of the form defined in \eqref{POVMproj}, we construct a noise POVM 
\begin{equation}
    \bm{N} = \left\{ N_1,...,N_k\right\},~~N_i = \lambda_i \left( \mathbb{1} - \ket{\phi_i}\!\!\bra{\phi_i} \right).
\end{equation}
Then 
\begin{equation}
    G[\bm{N}] = \sum_i \textrm{Tr}(A_i N_i) = \sum_i \lambda_i l_i^2 - p_i l_i^2 + 2 \dot p_i l_i = F_\textrm{C} + \sum_i \lambda_i l_i^2,
\end{equation}
and therefore
\begin{equation}
    \chi[\bm{M}, \bm{N}] = 2+ F_\textrm{Q}^{-1} \sum_i \lambda_i l_i^2,
\end{equation}
where we substituted $F_\textrm{C}$ with $F_\textrm{Q}$ because $\bm{M}$ is QCRB-sat. In our case, $l_i = - \sqrt{F_\textrm{Q}} \tan (\varphi_i/2)$, so
\begin{equation}
\label{bineq2m2}
    \chi[\bm{M}, \bm{N}] = 2 - \sum_i \lambda_i + \sum_i \lambda_i \cos^{-2}(\varphi_i/2) = 2 \sum_i \frac{\lambda_i}{2} \left| \cos^{-1} (\varphi_i/2) \right|^2,
\end{equation}
where we used the identity $\sum_i \lambda_i = \textrm{Tr}(\mathbb{1})=2$. Using power mean inequality (between powers $2$ and $-2$, $\lambda_i/2$ are weights), we obtain
\begin{equation}
    \label{ineq2m2}
    \sqrt{\sum_i \frac{\lambda_i}{2} \left| \cos^{-1} (\varphi_i/2)  \right|^2} \ge \left(\sum_i \frac{\lambda_i}{2} \left| \cos^{-1} (\varphi_i/2)  \right|^{-2} \right)^{-1/2} = \sqrt{2},
\end{equation}
the last equality follows from the fact that $\sum_i \lambda_i \cos^2(\varphi_i/2) = \frac{1}{4} \sum_i \lambda_i (2+e^{i\varphi_i} + e^{- i \varphi_i})=1$ because the sum over $i$ of anti-diagonal terms of $M_i$, which are $\lambda_i e^{\pm i \varphi_i}$, is equal to $0$. After inserting \eqref{ineq2m2} into \eqref{bineq2m2}, and using an inequality $\chi[\bm{M}] \ge \chi[\bm{M}, \bm{N}] $, we obtain that $\chi[\bm{M}] \ge 4$ for any QCRB-sat. $\bm{M}$. This inequality can only be saturated if \eqref{ineq2m2} is saturated, which means that all terms in the average are equal, and therefore $\cos^2(\varphi_1/2) = ... = \cos^2(\varphi_K/2)$. Then we have $1 = \sum_i \lambda_i \cos^2(\varphi_i/2) = \cos^2(\varphi_1/2) \sum_i \lambda_i  = 2 \cos^2(\varphi_1/2) $, so $\cos^2(\varphi_1/2) = ... = \cos^2(\varphi_K/2)=1/2$, and consequently $\varphi_i = \pm \pi/2$ for each $i$. Therefore, $\chi[\bm{M}]=4$ iff the measurement is a projective measurement on eigenstates of $\sigma_y$. 
\subsection{Super-resolution optical imaging}
\label{AppE2}
Let us firstly follow the construction from \cite{Tsang2016} of vectors $\ket{0}_\textrm{s}$,$\ket{1}_\textrm{s}$,$\ket{0}_\textrm{a}$,$\ket{0}_\textrm{a}$, where $\ket{0/1}_\textrm{s}$ form o.-n. basis of $\mathcal{H}_\textrm{s}$, $\ket{0/1}_\textrm{a}$ form o.-n. basis of $\mathcal{H}_\textrm{a}$, $\mathcal{H}^{(\theta)} = \mathcal{H}_\textrm{s} \oplus \mathcal{H}_\textrm{a}$, for a fixed value of $\theta$, $\rho_\theta, \dot \rho_\theta \in \mathcal{L} \left( \mathcal{H}^{(\theta)} \right)$. We define
\begin{align}
\label{basis1start}
    \ket{0}_\textrm{s} &= \frac{1}{\sqrt{2 (1+\delta)}} \left( \ket{u_{+,\theta}} + \ket{u_{-,\theta}} \right), \\
    \ket{1}_\textrm{s} &= \frac{1}{c_4} \left[ -\sqrt{2} \left(  \ket{\dot u_{+,\theta}}+\ket{ \dot u_{-,\theta}}  \right) + \frac{\gamma}{\sqrt{1+\delta}} \ket{0}_\textrm{s}\right], \\
    \ket{0}_\textrm{a} &= \frac{1}{\sqrt{2 (1-\delta)}} \left( \ket{u_{+,\theta}} - \ket{u_{-,\theta}} \right),\\
    \ket{1}_\textrm{a} &= \frac{1}{c_3} \left[ -\sqrt{2} \left(  \ket{\dot u_{+,\theta}}- \ket{\dot u_{-,\theta}}  \right) - \frac{\gamma}{\sqrt{1-\delta}} \ket{0}_\textrm{a}\right], 
    \label{basis1end}
\end{align}
where dot over vector denotes its derivative over $\theta$, and
\begin{align}
\label{coef1start}
    \delta &= \braket{u_{+,\theta}|u_{-,\theta}}= e^{-\theta^2/8 \sigma^2},\\
    \gamma &= 2 \braket{u_{+,\theta}|\dot u_{-,\theta}} = -\frac{\theta  e^{-\frac{\theta ^2}{8 \sigma ^2}}}{4 \sigma ^2}, \\
    c_3 &=\sqrt{ 4 \braket{ \dot u_{-,\theta}|\dot u_{-,\theta}} - 4 \braket{ \dot u_{+,\theta}|\dot u_{-,\theta}} - \frac{\gamma^2}{1-\delta}  }   = \frac{1}{4} \sqrt{\frac{8 \sigma ^2 \sinh \left(\frac{\theta ^2}{8 \sigma ^2}\right)-\theta ^2}{\sigma ^4 \left(e^{\frac{\theta ^2}{8 \sigma ^2}}-1\right)}},\\
    c_4 &= \sqrt{ 4 \braket{ \dot u_{-,\theta}|\dot u_{-,\theta}} + 4 \braket{ \dot u_{+,\theta}|\dot u_{-,\theta}} - \frac{\gamma^2}{1+\delta}  } = \frac{1}{4} \sqrt{\frac{8 \sigma ^2 \sinh \left(\frac{\theta ^2}{8 \sigma ^2}\right)+\theta ^2}{\sigma ^4 \left(e^{\frac{\theta ^2}{8 \sigma ^2}}+1\right)}}.
    \label{coef1stop}
\end{align}
Direct computations show, that $\ket{0}_\textrm{s}$, $\ket{1}_\textrm{s}$,$\ket{0}_\textrm{a}$,$\ket{1}_\textrm{a}$ indeed form o.-n. basis, and $\rho_\theta$, $\dot \rho_\theta$ can be expressed in this basis as
\begin{equation}
\label{matrices}
\rho_\theta = \frac{1}{2}\begin{bmatrix}1+\delta & 0 &0&0 \\
0 & 0 &0&0 \\
0 & 0 &1-\delta&0 \\
0 & 0 &0&0 \\
\end{bmatrix}, ~ \dot \rho_\theta = \frac{1}{2} \begin{bmatrix}\gamma & -\frac{c_4}{2} \sqrt{1+\delta} & 0 & 0 \\
 -\frac{c_4}{2} \sqrt{1+\delta} & 0  & 0 & 0 \\
0 & 0 & -\gamma &  -\frac{c_3}{2} \sqrt{1-\delta } \\
 0 & 0   &  -\frac{c_3}{2} \sqrt{1-\delta } & 0, \\
\end{bmatrix}
\end{equation}
which is equivalent to \eqref{rhosup}, \eqref{drhosup} after substituting
\begin{equation}
    \alpha = \frac{\gamma}{2},~ \beta_\textrm{s} = -\frac{c_4}{4} \sqrt{1+\delta},~ \beta_\textrm{a} = -\frac{c_3}{4} \sqrt{1-\delta }.
\end{equation}
Both $\rho_\theta$ and $\dot \rho_\theta$ have block-diagonal structure (as in Theorem 3):
\begin{equation}
    \rho_\theta = \rho_\theta^{(\textrm{s})} +\rho_\theta^{(\textrm{a})},~ \dot \rho_\theta =\dot \rho_\theta^{(\textrm{s})} +\dot \rho_\theta^{(\textrm{a})},  
\end{equation}
normalized components of $\rho_\theta$ are $\tilde \rho_\theta^{(\textrm{s/a})} = \ket{0}_\textrm{s/a}\!\!\bra{0}$,  their derivatives: $\dot{ \tilde{\rho}}_\theta^{(\textrm{s})}= \frac{2 \beta_{\textrm{s}}}{1+\delta} \sigma_x^{\textrm{(s)}} $, $\dot{ \tilde{\rho}}_\theta^{(\textrm{a})}= \frac{2 \beta_{\textrm{a}}}{1-\delta} \sigma_x^{\textrm{(a)}}$. Therefore, the evolution of the normalized density matrix in each subspace is locally equivalent to rotation of a pure state $\ket{0}$ around $y$ axis of a Bloch sphere. Consequently, taking into account Theorem 3 and the characterization of QCRB-sat. measurements for pure states from Section \ref{AppE1}, we conclude that potential candidates for minimal MeNoS QCRB-sat. measurements are of the form
\begin{equation}
    \bm{M} = \left\{\bm{M}^\textrm{(s)}, \bm{M}^\textrm{(a)}  \right\},~
    \bm{M}^\textrm{(s)} = \left\{  M_1^\textrm{(s)},...,M_{K_1}^\textrm{(s)} \right\},~\bm{M}^\textrm{(a)} = \left\{  M_1^\textrm{(a)},...,M_{K_2}^\textrm{(a)} \right\}
\end{equation}
where $M_i^\textrm{(s/a)} = \lambda_{i,\textrm{s/a}} \ket{\varphi_{i,\textrm{s/a}}}_\textrm{s/a} \!\!\bra{\varphi_{i,\textrm{s/a}}}$ , $ \ket{\varphi}_\textrm{s/a}=\cos(\varphi/2)\ket{0}_{s/a} + \sin(\varphi/2) \ket{1}_{s/a}$. In this case, QCRB-sat. measurements in each subspace consist on projectors on states from a meridian of a Bloch sphere, not from equator, because parametrization is different from the one used in Section \ref{AppE1}. As the last step, let us prove that $\bm{M}$ with a minimal MeNoS is of the form \eqref{supresoptM}, which means that $K_1=K_2=2$.  We will prove,that if $K_1 \ge 3$, then it is possible to construct a QCRB-sat. measurement $ \bm{ \tilde M} = \left\{ \bm{\tilde M}^\textrm{(s)}, \bm{ M}^\textrm{(a)} \right\}$ such that $\chi[\bm{\tilde M}] \le \chi[\bm{M}]$ and $\bm{\tilde M}^\textrm{(s)}$ contains $K_1-1$ elements. Let us start with choosing two indices $i,j \in \{1,...,K_1\}$ such that
\begin{equation}
\label{ijconds}
    \varphi_{i,\textrm{s}} \le \varphi_{j,\textrm{s}}~ \land~ \left( \varphi_{i,\textrm{s}},\varphi_{j,\textrm{s}} \in \left[0,\pi\right] \lor \varphi_{i,\textrm{s}},\varphi_{j,\textrm{s}} \in \left[\pi,2\pi\right]  \right).
\end{equation} Such a choice is always possible for $K_1 \ge 3$ because of pigeonhole principle. After some algebra, we obtain the following identity
\begin{equation}
\label{MiMjsum}
    M_i^\textrm{(s)} + M_j^\textrm{(s)} = A \mathbb{1}_\textrm{s} + B \ket{\tilde \varphi}_\textrm{s}\!\!\bra{\tilde \varphi},
\end{equation}
where
\begin{equation}
    B=\sqrt{\lambda_{i,\textrm{s}}^2+\lambda _{j,\textrm{s}}^2+2 \lambda_{i,\textrm{s}} \lambda_{j,\textrm{s}} \cos \left(\varphi_{j,\textrm{s}}-\varphi_{i,\textrm{s}} \right)},~~ A= \frac{1}{2} \left(\lambda_{i,\textrm{s}}+\lambda_{j,\textrm{s}}-B\right)
\end{equation}
are positive constants, $A < 1$ because $\lambda_{i,\textrm{s}}+\lambda_{j,\textrm{s}} < \textrm{Tr} (\mathbb{1}_\textrm{s}) = 2$, and
\begin{equation}
    \tan\left( \tilde \varphi \right) = \frac{C \tan \left(\varphi _{i,\textrm{s}}\right)+D \tan \left(\varphi _{j,\textrm{s}}\right)}{C+D},~~C=\lambda _{i,\textrm{s}} \cos^{-1} \left(\varphi _{j,\textrm{s}}\right),~~D=\lambda _{j,\textrm{s}} \cos^{-1} \left(\varphi _{i,\textrm{s}}\right).
\end{equation}
From the above formula it is clear that  $\tan(\tilde \varphi)$ lies between $\tan(\varphi_{i,\textrm{s}})$ and $\tan(\varphi_{j,\textrm{s}})$, so using \eqref{ijconds}, we obtain $\tilde \varphi \in [\varphi_{i,\textrm{s}},\varphi_{j,\textrm{s}} ]$. After removing  $M_i^\textrm{(s)}$ and $M_j^\textrm{(s)}$ from $\bm{M}^\textrm{(s)}$, and replacing them with $B \ket{\tilde \varphi}_\textrm{s}\!\!\bra{\tilde \varphi}$, the sum of all elements is decreased from $\mathbb{1}_\textrm{s}$ to $(1-A) \mathbb{1}_\textrm{s} $, according to \eqref{MiMjsum}. Therefore, to obtain a valid POVM we need to rescale all elements after such a replacement, and then we end up with
\begin{equation}
    \bm{\tilde M}^\textrm{(s)} =  \left\{ \left.(1-A)^{-1} M_k^\textrm{(s)} \right| _{k \in \{1,...,K_1 \} \setminus \{i,j\} } , (1-A)^{-1} B \ket{\tilde \varphi}_\textrm{s}\!\!\bra{\tilde \varphi}  \right\},
\end{equation}
which is a POVM containing $K_1-1$ elements. From the construction, $\bm{\tilde M}^\textrm{(s)}$ contains only elements proportional to projectors on a Bloch sphere meridian, so it is QCRB-sat. Let us now prove, that for a given construction, $\chi[\bm{\tilde M}] \le \chi[\bm{M}]$.
The logarithmic derivative of $p_\theta(k,\textrm{s}) = \textrm{Tr} \left( \rho_\theta M_k^{\textrm{(s)}} \right)$ is
\begin{equation}
    l_{k,\textrm{s}} = \frac{\textrm{Tr} \left( \dot \rho_\theta M_k^{\textrm{(s)}} \right)}{\textrm{Tr} \left( \rho_\theta M_k^{\textrm{(s)}} \right)} = f(\varphi_{k,\textrm{s}}),~\textrm{where}~f(\varphi) = \frac{2}{1+\delta} \left( \alpha + 2 \beta_\textrm{s} \tan (\varphi/2) \right), 
\end{equation}
notice that $l_{k, \textrm{s}}$ does not depend on $\lambda_{k,\textrm{s}}$, and that $f(\varphi)$ is monotonic in range $\left[0,\pi\right]$ and in range $\left[\pi,2\pi\right]$. That means, that after replacing $\bm{M}^\textrm{(s)}$ with $\bm{\tilde M}^\textrm{(s)}$, logarithmic derivatives of outcomes with indices $k \in \{1,...,K_1 \} \setminus \{i,j\}$ are not affected, and  $l_{i, \textrm{s}}$, $l_{j, \textrm{s}}$ are replaced with a logarithmic derivative $\tilde l = f(\tilde \varphi)$. From inclusion $\tilde \varphi \in [\varphi_{i,\textrm{s}},\varphi_{j,\textrm{s}} ]$, we conclude that $\tilde l$ lies between $l_{i, \textrm{s}}$ and $l_{j, \textrm{s}}$. This means, that the described modification of $\bm{M}$ neither increased the largest logarithmic derivative, nor decreased the smallest one. Therefore, according to Theorem 4, $\chi[\bm{\tilde M}] \le \chi[\bm{M}]$. It means, that we can lower the value of $K_1$ without affecting the CFI, and without increasing MeNoS, as long as $K_1>2$. The same reasoning can be applied to the assymetric subspace, so we can also lower $K_2$. Therefore, we can always find a QCRB-sat. measurement with a minimal MeNoS with $K_1=K_2=2$, and such a measurement must be of the form defined in \eqref{supresoptM}.

In the main text, we also considered the family of measurements in Hermite-Gaussian modes, $M_i= \ket{\phi_i}\!\!\bra{\phi_i}$ for $i \in \{1,2,...,K-1\}$, $M_K = \mathbb{1} - M_1-...-M_{K-1}$, where the representation of $\ket{\phi_i}$ in a position basis is
\begin{equation}
    \braket{x|\phi_i}= \left( \frac{1}{2 \pi \sigma^2} \right)^{1/4} \frac{1}{\sqrt{2^i i!}} H_q \left( \frac{x}{\sqrt 2 \sigma} \right) \exp \left( - \frac{x^2}{4 \sigma^2} \right).
\end{equation}
According to Ref.\cite{Tsang2016}, the probability of obtaining $i$-th outcome for an input state defined in \eqref{imagingstate} is
\begin{equation}
    p_\theta(i) = \braket{\phi_i|\rho_\theta|\phi_i} = \exp \left(- \theta^2/16\sigma^2 \right) \frac{\left( \theta^2/16\sigma^2 \right)^i}{i!}~~\textrm{for}~i \in \{1,...,K-1\},~~ p_\theta(K) = 1-p_1-...-p_{K-1}, 
\end{equation}
 corresponding logarithmic derivatives are
\begin{equation}
    l_i = -\frac{\theta}{8 \sigma^2} + \frac{2i}{\theta} ~~\textrm{for}~i \in \{1,...,K-1\},~~l_K = -\frac{l_1 p_1 + ... + l_{K-1} p_{K-1}}{p_K}.
\end{equation}
Both CFI and FI MeNoS plotted in Fig.~\ref{Fig:superresolution} are computed directly using above relations and formulas \eqref{CFIdef} and \eqref{chian}.
\section{The role of correlations in super-resolution imaging}
\label{AppF}
For some types of sources, subsequent photon emissions are not independent, and then many-photon density matrix is not separable, i.e. $\rho_\theta^{(N)} \ne \left(\rho_\theta^{(1 )} \right)^{\otimes N} $, where $\rho_\theta^{(1)}$ is defined in \eqref{imagingstate}. Then, the CFI and QFI for $\rho_\theta^{(N)}$ are not necessarily $N$ times larger than for $\rho_\theta^{(1)}$, so the results obtained using the one-photon model cannot be applied directly. 
It was shown, that correlations between subsequent photons, resulting from super-bunching or anti-bunching phenomena, can be used to increase the imaging resolution within the \textit{direct imaging} paradigm, where each photon in the image plane is measured in the position basis \cite{dertinger2009fast, kurdzialek2021super, schwartz2012improved}. 

It is natural to ask, if the advantage from correlations can be also seen if we do not restrict ourselves to direct imaging, but rather assume, that any quantum measurement can be performed. To get some intuition regarding this problem, let us consider a simple toy-model involving correlations---we assume, that sources always emit photons in pairs and the time between subsequent pairs is much longer than the time between emissions of two photons within one pair. Consequently, each pair comes from one source, but we do not know from which one, and the state of a photon pair is
\begin{equation}
    \tilde \rho_\theta^{(2)} =  \frac{1}{2} \left( \ket{u^{(2)}_{+,\theta}}\!\!\bra{u^{(2)}_{+,\theta}} + \ket{u^{(2)}_{-,\theta}}\!\!\bra{u^{(2)}_{-,\theta}}  \right),
\end{equation}
where $\ket{u^{(2)}_{\pm,\theta}} = \ket{u_{\pm,\theta}}^{\otimes 2} 
$. This density matrix has the same structure as $\rho_\theta^{(1)}$, so we can construct 4-element o.-n. basis spanning $\tilde \rho_\theta^{(2)}$ and $\partial_\theta \tilde \rho_\theta^{(2)}$, as in (\ref{basis1start}--\ref{basis1end}):  
\begin{align}
\label{basis2start}
     \ket{\tilde 0}_\textrm{s} &= \frac{1}{\sqrt{2 (1+\tilde \delta)}} \left( \ket{u^{(2)}_{+,\theta}} + \ket{u^{(2)}_{-,\theta}} \right), \\
    \ket{\tilde 1}_\textrm{s} &= \frac{1}{\tilde c_4} \left[ -\sqrt{2} \left(  \ket{\dot u^{(2)}_{+,\theta}}+\ket{ \dot u^{(2)}_{-,\theta}}  \right) + \frac{\tilde \gamma}{\sqrt{1+\tilde \delta}} \ket{\tilde 0}_\textrm{s}\right], \\
    \ket{\tilde 0}_\textrm{a} &= \frac{1}{\sqrt{2 (1-\tilde \delta)}} \left( \ket{u^{(2)}_{+,\theta}} - \ket{u^{(2)}_{-,\theta}} \right),\\
    \ket{\tilde 1}_\textrm{a} &= \frac{1}{\tilde c_3} \left[ -\sqrt{2} \left(  \ket{\dot u^{(2)}_{+,\theta}}- \ket{\dot u^{(2)}_{-,\theta}}  \right) - \frac{\tilde \gamma}{\sqrt{1-\tilde \delta}} \ket{\tilde 0}_\textrm{a}\right], 
    \label{basis2end}
\end{align}
where
\begin{align}
\label{coef2start}
    \tilde \delta &= \braket{u^{(2)}_{+,\theta}|u^{(2)}_{-,\theta}}= e^{-\theta^2/4 \sigma^2},\\
    \tilde \gamma &= 2 \braket{u^{(2)}_{+,\theta}|\dot u^{(2)}_{-,\theta}} = -\frac{\theta  e^{-\frac{\theta ^2}{4 \sigma ^2}}}{2 \sigma ^2}, \\
    \tilde c_3 &=\sqrt{ 4 \braket{ \dot u^{(2)}_{-,\theta}|\dot u^{(2)}_{-,\theta}} - 4 \braket{ \dot u^{(2)}_{+,\theta}|\dot u^{(2)}_{-,\theta}} - \frac{\tilde \gamma^2}{1-\tilde \delta}  }   = \frac{1}{2} \sqrt{\frac{4 \sigma ^2 \sinh \left(\frac{\theta ^2}{4 \sigma ^2}\right)-\theta ^2}{\sigma ^4 \left(e^{\frac{\theta ^2}{4 \sigma ^2}}-1\right)}},\\
    \tilde c_4 &= \sqrt{ 4 \braket{ \dot u^{(2)}_{-,\theta}|\dot u^{(2)}_{-,\theta}} + 4 \braket{ \dot u^{(2)}_{+,\theta}|\dot u^{(2)}_{-,\theta}} - \frac{\tilde \gamma^2}{1+\tilde \delta}  } = \frac{1}{2} \sqrt{\frac{4 \sigma ^2 \sinh \left(\frac{\theta ^2}{4 \sigma ^2}\right)+\theta ^2}{\sigma ^4 \left(e^{\frac{\theta ^2}{4 \sigma ^2}}+1\right)}}.
    \label{coef2stop}
\end{align}
The expressions for $\tilde \rho_\theta^{(2)}$ and $\partial_\theta \tilde \rho_\theta^{(2)}$ in the introduced basis are the same as the expressions for $\rho_\theta^{(1)}$ and $\dot \rho_\theta^{(1)}$ \eqref{matrices}, the only difference is that $\delta$, $\gamma$, $c_3$, $c_4$ are replaced with $\tilde \delta$, $\tilde \gamma$, $\tilde c_3$, $\tilde c_4$. After comparing Eqs. (\ref{coef1start}--\ref{coef1stop}) with Eqs. (\ref{coef2start}--\ref{coef2stop}) we see, that formulas for $\tilde \rho_\theta^{(2)}$ and $\partial_\theta \tilde \rho_\theta^{(2)}$ are obtained by replacing $\sigma$ with $\sigma/\sqrt{2}$ in formulas for $\rho_\theta^{(1)}$ and $\dot \rho_\theta^{(1)}$. Therefore, two-photon model described by $\tilde \rho_\theta^{(2)}$ is equivalent to one-photon model $\rho_\theta^{(1)}$ with $\sigma$ narrowed by a factor $\sqrt{2}$. Using this observation, and the fact that $F_\textrm{Q}[\rho_\theta^{(1)}]= \frac{1}{4 \sigma^2}$, we obtain
\begin{equation}
\label{QFIcor}
    F_\textrm{Q}[\tilde \rho_\theta^{(2)}] = \frac{1}{4 (\sigma/\sqrt{2})^2} = 2 F_\textrm{Q} [ \rho_\theta^{(1)}]
\end{equation}

The density matrix describing two uncorrelated photons is $\rho_\theta^{(2)} = \rho_\theta^{(1)} \otimes \rho_\theta^{(1)}$, and from additivity of QFI, we have $F_\textrm{Q}[\rho_\theta^{(2)}] = 2 F_\textrm{Q} [ \rho_\theta^{(1)}] = F_\textrm{Q} [\tilde \rho_\theta^{(2)}]$. This may suggest, that super-bunching based correlations of the considered type do not provide any increase of the imaging precision, when all quantum measurements are allowed. However, this is only true for a noiseless scenario---let us check, how small measurement noise affects the esimation precision in both cases---with and without correlations.

For the correlated model described by $\tilde \rho_\theta^{(2)}$, we consider arbitrary (possibly correlated) measurement $\bm{M}^{(2)}$, which can be affected by arbitrary noise $\bm{N}^{(2)}$, such that noise-affected two-photon POVM is $\bm{\tilde M}^{(2)} = (1-\epsilon)\bm{M}^{(2)} + \epsilon \bm{N}^{(2)}$, and then the resulting noise-affected CFI is, up to terms of the order of $\epsilon$,
\begin{equation}
    F_\textrm{C}[\tilde \rho_\theta^{(2)}, \bm{\tilde M}^{(2)}] = F_\textrm{C}[\tilde \rho_\theta^{(2)}, \bm{ M}^{(2)}]\left( 1 - \epsilon \chi[\bm{M}^{(2)}, \bm{N}^{(2)}] \right).
\end{equation}
The lowest possible FI MeNoS of QCRB-sat. measurement for this model is
\begin{equation}
    \tilde \chi_\textrm{Q}^{(2)} \left( \frac{\theta}{\sigma} \right) = \underset{\left\{\bm{M}^{(2)} \in \mathcal{M},  F_\textrm{C}\left[ \bm{M}^{(2)} \right] = F_\textrm{Q}\right\}}{\textrm{min}} \chi \left[ \bm{M}^{(2)} \right].
\end{equation}
Fortunately, we do not need to repeat the whole procedure from Section \ref{AppE2} to calculate $\tilde \chi_\textrm{Q}^{(2)} \left( \frac{\theta}{\sigma} \right)$ because models described by $\rho_\theta^{(1)}$ and $\tilde \rho_\theta^{(2)}$ are isomorphic, and the only difference between them is that parameter $\sigma$ is decreased by a factor $\sqrt{2}$ in the latter one. Consequently,
\begin{equation}
    \tilde \chi_\textrm{Q}^{(2)} \left( \frac{\theta}{\sigma} \right) = \chi_\textrm{Q}^{(1)} \left( \frac{\theta}{\sigma \sqrt{2}} \right),
\end{equation}
where $\chi_\textrm{Q}^{(1)}$ is $\chi_\textrm{Q}$ computed for $\rho_\theta^{(1)}$. 

\begin{figure}[h]
\includegraphics[width=10cm]{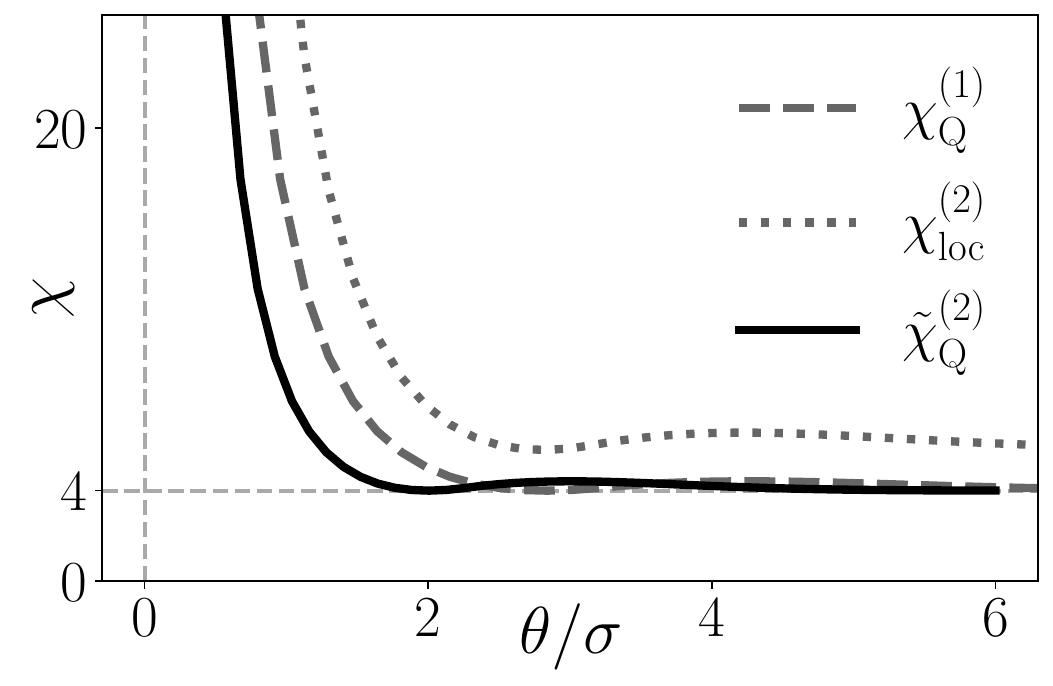}
\caption{Two-photon correlations allow to achieve a lower noise susceptibility ($\tilde \chi_\textrm{Q}^{(2)}$) than for non-correlated case ($\chi_\textrm{loc}^{(2)}$). Optimal one-photon noise susceptibility ($\chi_\textrm{Q}^{(1)}$) can be interpreted as two-photon noise susceptibility for non-correlated case, when we restrict ourselves to non-correlated noise.  }
\label{Figapp}
\end{figure}

Let us now examine the scenario with uncorrelated photons ($\rho^{(2)}_\theta = \rho^{(1)}_\theta \otimes \rho^{(1)}_\theta$) and uncorrelated one-photon measurements ($\bm{M}^{(2)} = \bm{M}^{(1)} \otimes \bm{M}^{(1)} $). We choose $\bm{M}^{(1)}$ to be the lowest susceptibility QCRB-sat. measurement for $\rho_\theta^{(1)}$, found in Section \ref{AppE2}, and compute $\chi^{(2)}_\textrm{loc} = \chi[\bm{M}^{(1)} \otimes \bm{M}^{(1)}]$ directly from \eqref{chian}. As we see in Fig.~\ref{Figapp}, $\chi_\textrm{loc}^{(2)} > \chi^{(1)}$. This is because  $\chi[\bm{M}^{(1)} \otimes \bm{M}^{(1)}]$ is computed by maximizing $\chi[\bm{M}^{(1)} \otimes \bm{M}^{(1)}, \bm{N}^{(2)}]$ over all noise POVMs $\bm{N}^{(2)}$, also non-local ones. Therefore, the maximization is taken over a larger class of noise than for one-photon model. 

It is often reasonable to assume, that for uncorrelated measurements on uncorrelated systems, the same noise affects each one-photon POVM independently, such that $\bm{M}^{(1)}$ becomes $\bm{\tilde M}^{(1)} = (1- \epsilon) \bm{M}^{(1)} + \epsilon \bm{N}^{(1)}$, and consequently $\bm{M}^{(2)}$ becomes $\bm{\tilde M}^{(2)} = \bm{\tilde M}^{(1)} \otimes \bm{\tilde M}^{(1)}$. Then, the noise affected CFI is
\begin{equation}
    F_\textrm{C}[\rho_\theta^{(2)}, \bm{\tilde M}^{(2)} ] = 2 F_\textrm{C} [\rho_\theta^{(1)}, \bm{\tilde M}^{(1)} ] = 2 F_\textrm{C}[\rho_\theta^{(1)}, \bm{M}^{(1)}] \left(1 - \epsilon \chi[\bm{M}^{(1)}, \bm{N}^{(1)}] \right) = F_\textrm{C}[\rho_\theta^{(2)}, \bm{M}^{(2)}] \left(1-\epsilon \chi[\bm{M}^{(1)}, \bm{N}^{(1)}] \right),
\end{equation}
terms of the order of $\epsilon^2$ were omitted. When $\bm{M}^{(1)}$ is the lowest susceptibility QCRB-sat. measurement, then
\begin{equation}
    F_\textrm{C}[\rho_\theta^{(2)}, \bm{\tilde M}^{(1)} \otimes \bm{\tilde M}^{(1)} ] \ge F_\textrm{Q}[\rho^{(2)}] (1-\epsilon \chi_\textrm{Q}^{(1)})
\end{equation}
We see, that the optimal noise susceptibility for two non-correlated photons is $\chi_\textrm{Q}^{(1)}$, assuming non-correlated QCRB-sat. measurement and non-correlated noise. Notice, that when $\epsilon$-small noise $\bm{N}^{(1)}$ acts on each photon, then two photons are affected by $2 \epsilon$-small noise because
\begin{equation}
    \bm{\tilde M}^{(1)} \otimes \bm{\tilde M}^{(1)} = (1- 2 \epsilon) \bm{M}^{(1)} \otimes \bm{M}^{(1)} + 2 \epsilon \bm{N}^{(2)}, 
\end{equation}
where $\bm{N}^{(2)} = \frac{1}{2} (\bm{M}^{(1)} \otimes \bm{N}^{(1)} + \bm{N}^{(1)} \otimes \bm{M}^{(1)})$ is a two-photon noise POVM. However, from comparison between $\chi_\textrm{Q}^{(1)}$ and $\chi_\textrm{loc}^{(2)}$ follows, that the effect of non-correlated $2 \epsilon$-noise is weaker than the effect of $\epsilon$-small correlated noise.

As we see in Fig.~\ref{Figapp}, we have $\tilde \chi_\textrm{Q}^{(2)}<\chi_\textrm{loc}^{(2)}$. Consequently, even though correlations do not increase the QFI, they allow to obtain better robustness against noise, and consequently better precision of the estimation of $\theta$ in a noisy environment. The advantage for $\theta<2 \sigma$ is present even if we assume that for the non-correlated model, noise also has to be non-correlated, because then $\tilde \chi_\textrm{Q}^{(2)}<\chi^{(1)}_\textrm{Q}$.

To sum up, we considered a simple model of two-photon correlations, and demonstrated their usefulness in achieving better precision of the separation estimation between two sources in the presence of noise. Further research is required to study more realistic correlation models, and search for practical, noise-robust protocols utilizing correlations between photons and correlated measurements. 

\end{document}